\documentclass[showpacs,preprintnumbers,amssymb,twocolumn]{revtex4-2}

\usepackage{graphicx}
\usepackage{dcolumn}
\usepackage[dvipsnames]{xcolor}
\usepackage{bm}
\usepackage{amsmath}
\usepackage{amssymb}
\usepackage{epsfig}
\usepackage{amsfonts}
\usepackage{lineno,hyperref}
\usepackage{array}
\usepackage{float}
\usepackage{microtype}
\usepackage{multirow}
\usepackage{adjustbox}
\usepackage[english]{babel}
\usepackage{epstopdf}
\usepackage{blindtext}
\usepackage{subcaption}
\usepackage[a4paper, total={6.7in, 10in}]{geometry}
\usepackage{appendix}
\usepackage{xcolor}

\def \l{\lambda}

\def \h{\hbar}
\def \g{\gamma}

\def \O{\Omega}

\def \s{\sqrt}
\def \D{\Delta}

\def \be{\begin{equation}}
\def \ee{\end{equation}}
\def \ben{\begin{eqnarray}}
\def \een{\end{eqnarray}}

\def \G{\Gamma}

\def \n{\nonumber}
\def \N{\aleph}

\begin{document}
\title{Connecting Gravity and Quantum Physics: Primordial Black Holes and Accelerated Evolution of the Universe}

\author{ Victor Borsevici}
\email{borsevici@mail.ru}
\affiliation{ Universitatea Libera Internationala din Moldova, International Informatization Academy, a branch of Republic of Moldova,\\
 Chisinau, Republic of Moldova}

\author{Samit Ganguly}
\email{samitganguly1994@gmail.com}
\affiliation{Department of Physics, Haldia Government College, Haldia, Purba Medinipur 721657, India $\&$ \\ Department of Physics, University of Calcutta, 92, A.P.C. Road, Kolkata-700009, India}

\author{Goutam Manna$^a$}
\email{goutammanna.pkc@gmail.com \\$^a$Corresponding author}
\affiliation{Department of Physics, Prabhat Kumar College, Contai, Purba Medinipur 721404, India $\&$\\ Institute of Astronomy Space and Earth Science, Kolkata 700054, India}

\date{\today}

\begin{abstract}
This study presents a new framework to explore the fundamental relationship between gravity and quantum mechanics, with particular emphasis on the fundamental role of primordial black holes (PBHs) in cosmology. Through the concept of self-gravitating condensed light in the form of the experimentally discovered quantum photon Bose-Einstein condensate, this work examines the quantized gravitational, informational, thermodynamical, traditional, and other attributes of PBHs and their implications for early universe dynamics, baryogenesis, the very early formation of galaxies, supermassive black holes (SMBHs) and large-scale structures. Precise calculations have shown that primordial protogalaxies with supermassive black holes at their centers of gravity were formed before the recombination epoch. By solving issues like the cosmological constant problem and the information loss paradox, dark matter and dark energy, this work provides insights into Planck-scale physics and it's impact to cosmology. In such a way PBHs serve as a bridge between quantum theory and general relativity. This study ultimately posits that presented PBH physics is essential to resolving major cosmological and astrophysical issues, paradoxes and "mysteries", such as the accelerated evolution of the Universe established by JWST and other observations.

\end{abstract}

%\keywords{Primordial black holes, Bose-Einstein Condensate, Planck Scale Physics, Early Universe }
%%%%%%%%%%%%
%\pacs{03.65.−w, 03.65.Fd, 03.75.Nt, 04.60.−m, 04.70.Dy}
%%%%%%%%%%%%%
\maketitle

%\keywords{Primordial black holes, Bose-Einstein Condensate, Planck Scale Physics, Dark matter and Dark energy, Early universe and Large-Scale Structures.}

%\ccode{PACS numbers: 04.20.-q, 04.20.Cv, 04.70.−s, 04.70.Dy}

%\tableofcontents

\section{Introduction}

The rapid advancement of modern experimental, observational, and measurement technologies—delivering precise and often unexpected data, as exemplified by the JWST—revives longstanding, unresolved questions in theoretical physics and cosmology. At the heart of these open problems lies the fundamental, yet elusive, connection between General Relativity and Quantum Theory.

The main goal of our work is to find and mathematically express this connection, and, on it's basis, to resolve or remove such major issues, paradoxes and "mysteries" as cosmological constant problem, informational loss paradox, cosmological principle problem related to the large-scale structure "excesses", physical nature and origin of dark matter and dark energy, baryogenesis, and baryonic assymetry, the very early growth of SMBHs as the gravitational centers of the very early proto-galaxies, the very early clumping of the Universe and it's accelerated evolution from the Beginning of Time. 

Since S. Hawking's time more and more cosmologists have begun to recognize the exceptional role of primordial black holes in cosmogenesis and cosmic evolution, but no one knew what kind of gravitating matter they were made of and what their real physical nature is. It should also be noted that since the time of J.Beckenstein, V.Mukhanov, and G.t'Hooft many people have also begun to talk about the quantum nature of black holes, comparing it with the quantum nature of hydrogen atom.

Fortunately, the recent experimental discovery of quantized condensed light predicted a century ago by A. Einstein in the form of photon Bose-Einstein condensate (photon BEC) gives us a unique opportunity to relate its Compton wavelength to the gravitational radius of a primordial black hole. This geometrical approach makes PBHs a unique bridge between quantum theory and general relativity.

It is very important to note that photons are the only particles whose wavelength is capable to cover the immense range of black hole sizes, ranging from Planck-scale sizes to tens of solar system diameters. This is what makes condensed light with its effective rest energy and effective rest mass the only candidate for the "substance" that black holes are made of.

It is the quantum nature of gravitating condensed light that predicts the quantized characteristics of the PBHs such as mass, amount of quantum information, entropy, temperature, radiation, quantum absorption, lifetime, etc. Also, it provides the simple and clear laws that govern their birth and death, their development, and their decisive role in the accelerated evolution of the Universe. Moreover, presented PBHs physics opens the door to the Terra incognita of Planck-scale physics and the beginning of time.

It should be especially noted that in this work we assume that modern dark matter consists of primordial black holes with a typical mass of 1.659x10$^{18}$gr and a minimum of 2.598x10$^{17}$gr, while dark energy is represented by the gravitational radiation associated with Dark Matter mass-energy loss processes, starting from the hot Big Bang to the present time.

%{\color{red}Moreover, the precise calculations presented in this paper show that the first protogalaxies with supermassive black holes at their centers, consisting of ordinary baryonic and dark matter, appeared before the recombination epoch. Let us add that at this time the Universe was a early universal clumped structure, which can be called a "universal supercluster", which, as the Universe expanded, "cracked" to form modern giant supervoids, filaments and superclusters.- rewrite it elegantly}

Moreover, the precise calculations presented in this paper indicate that the first protogalaxies, harboring supermassive black holes and composed of both ordinary baryonic and dark matter, emerged before the epoch of recombination. At this stage, the Universe existed as a primordial, clumped structure—a "universal supercluster." As cosmic expansion progressed, this vast structure fragmented, giving rise to the immense supervoids, filaments, and superclusters that define the large-scale cosmic web we observe today.

Let us introduce a brief look at a number of theoretical and practical studies that are related to the topic of our research.

Gravity and quantum mechanics are widely regarded as the two pillars of modern physics, each describing the universe with unparalleled success but in fundamentally different realms. Gravity, as encapsulated by Einstein's theory of General Relativity, governs the macroscopic world: the motion of planets, the dynamics of galaxies, and the curvature of spacetime itself. Its elegance lies in the geometric framework, where massive objects dictate the shape of spacetime, and spacetime, in turn, guides the motion of these objects.

Quantum mechanics, on the other hand, is the language of the microscopic world. It describes the behavior of particles at the atomic and subatomic scales governed by probabilities, wave functions, and inherent uncertainty. This framework has given rise to astonishing technologies and profound insights into the nature of matter and energy, yet it operates in a domain where classical notions of reality are often defied.

Despite their individual successes, gravity and quantum mechanics resist unification. General relativity treats spacetime as a smooth continuum, while quantum mechanics thrives on discreteness and probabilistic events. Attempts to bridge these paradigms often encounter insurmountable challenges: mathematical inconsistencies, infinite quantities, and a lack of experimental evidence at the scales where their interplay would manifest.

This profound incompatibility underscores the uniqueness of both frameworks and raises one of the most profound questions in physics: How can we reconcile the smoothness of spacetime with the granularity of quantum mechanics? The pursuit of a quantum theory of gravity, whether through string theory, loop quantum gravity, or other approaches, continues to push the boundaries of our understanding. It is not merely a quest for unification, but a journey to uncover a deeper, more fundamental description of reality.

The need for a quantum theory of gravity arises from the limitations of our current theoretical frameworks when applied to extreme conditions in the universe. General Relativity successfully describes the behavior of spacetime and gravity on large scales, but it breaks down under extreme conditions, such as within black holes or at the very beginning of the universe during the Big Bang. At these scales, quantum effects become significant, and a theory that combines both quantum mechanics and gravity is essential to provide a complete description of nature.

Without a quantum theory of gravity, our understanding of phenomena like the singularities in black holes or the quantum fluctuations that seeded the structure of the universe remains incomplete. Additionally, the unification of gravity with the other fundamental forces—electromagnetic, weak, and strong—requires a consistent framework that integrates all forces under the same quantum principles. A quantum theory of gravity would also have profound implications for the nature of spacetime, that helps revealing its granular structure and resolving paradoxes like the information loss problem in black holes.

The development of a quantum gravity theory is not only a theoretical pursuit but also a step toward addressing practical questions in cosmology, high-energy physics, and even quantum computing. It promises to bridge the gap between the macroscopic and microscopic and provides a unified view of the universe from the smallest particles to the vast cosmos.

In this regard, we want to mention that, the study of primordial black holes (PBHs) and two-dimensional (2D) photon Bose-Einstein condensate (BEC) are crucial for connecting gravity with quantum mechanics, as they naturally combine extreme gravitational and quantum mechanical effects. Formed shortly after the Big Bang, PBHs probe the interplay of quantum fluctuations and gravitational collapse under early-universe conditions, which offers unique insights into Planck-scale physics. Their potential evaporation via Hawking radiation provides a direct test of quantum field theory in curved spacetime, while their small masses allow us to validate the quantum gravity phenomena like black hole thermodynamics, information loss paradoxes, and potential remnants. PBHs also constrain cosmological models, link to dark matter hypotheses, and contribute to gravitational wave generation \cite{Green, Choudhury}. By studying PBHs, we can investigate quantum gravity theories, explore the statistical nature of black hole entropy, and seek observable signatures like gamma-ray bursts or microlensing, making them a potential candidate that may help us to unify quantum mechanics and general relativity\cite{Khlopov}.

Along with this, the recent developments in PBHs research have shown its substantial cosmic consequences, as emphasized by Carr et al. \cite{carr,Carr}. Their research offers convincing indirect evidence that PBHs, especially those with extensive mass distribution, could illustrate several astrophysical anomalies, such as the emergence of ultra-faint dwarf galaxies, the dynamical heating of the Galactic disk, and high-redshift quasars, presenting alternatives to the conventional cold dark matter paradigm. This corresponds with the increasing recognition of PBHs as significant factors in early cosmology, perhaps acting as progenitors for supermassive black holes (SMBHs) and facilitating the construction of cosmic structures. The authors of \cite{carr} present a ``positivist perspective," highlighting the unifying potential of PBHs in resolving outstanding astrophysical queries, further substantiated by new associations with photon Bose-Einstein condensates. Comprehensive research on PBHs and their cosmic significance, including their birth and growth, is documented in the following Refs. \cite{Khlopov1,Khlopov2, Belotsky1,Belotsky2,Hawking1}

On the other hand, BEC comprising a two-dimensional condensate photon gas \cite{Muller,Ruffini} offers a distinctive platform for investigating and evaluating ideas in quantum gravity, functioning as a quantum-coherent system that reflects characteristics of spacetime and gravitational processes. In such a condensate, the alteration of the effective potential encountered by photons can replicate curved spacetime, generating analogs of event horizons and gravity wells\cite{Klaers}. This facilitates the examination of phenomena such as Hawking radiation inside a regulated laboratory setting \cite{Belgiorno} under analog gravity, providing experimental insights into quantum field theory in curved spacetime\cite{Maldacena}.

The macroscopic coherence of the photon BEC also embodies characteristics fundamental to quantum gravity, facilitating the investigation of how quantum phenomena may influence the formation of spacetime and gravity \cite{Maldacena}. Quantum fluctuations, especially evident in two-dimensional systems, offer additional avenues for exploring dynamics similar to those of the early universe\cite{Schley,Bengochea}.   The thermal and entropic characteristics of the condensate align with the laws of black hole thermodynamics, providing an analogy to the Bekenstein-Hawking entropy \cite{Bekenstein1,Hawking} and enhancing our comprehension of quantum geometry.

Through the analysis of these systems, we can investigate the interaction between quantum mechanics and gravity, simulate emergent spacetime, and evaluate notions such as coherence, thermodynamics, and quantum field behavior. The 2D photon BEC connects theoretical physics with experimental practices, and may offer a profound opportunity to illuminate the path towards establishing a connection between quantum mechanics and general relativity.

Thus, the concept of a 2D photon BEC introduces a new approach to study the quantum statistical behavior of light by combining quantum optics, condensed matter physics, and thermodynamics. Unlike traditional atomic BECs, where atoms condense at low temperatures, photons—typically massless—can achieve a condensed state by simulating an effective mass within a planar optical microcavity. This is achieved by confining photons between curved mirrors, creating a 2D environment where they behave similarly to massive particles. A heat reservoir, often via dye molecules, allows the photons to reach thermal equilibrium, forming a "rest energy" and enabling condensation at reduced temperatures \cite{Klaers, Vretenar}. As the system cools, photons transition to the lowest energy mode, forming a coherent quantum state \cite{Kirton}. This 2D photon BEC platform offers significant insights into fundamental physics, potentially impacting low-threshold laser technology, quantum information, and the study of thermodynamic properties in non-equilibrium quantum systems \cite{Vretenar}. Recently, Mazur and Mottola \cite{Mazur} introduced the concept of Gravitational Condensate Stars as a stable alternative to black holes. In their work, they present a compact, non-singular solution to Einstein’s equations, suggesting it as a possible endpoint for gravitational collapse. Unlike black holes, this type of collapsed star aligns with quantum theory, remains thermodynamically stable, and avoids the information loss paradox.

Besides this, Recent remarkable discoveries, including Einstein’s gravitational waves \cite{GW1, GW2, GW3, GW4}, ``impossible" early galaxies and quasars \cite{Naidu, Napolitano, Labbe, Wang}, and the 2D photon Bose-Einstein condensate with rest energy \cite{Muller, Klaers,  Vretenar, Kirton}, may offer a strong experimental and observational foundation for the existence of a quantum gravity theory , which John Wheeler famously described as the ``fiery marriage of general relativity with quantum theory."

In this study, the objective is to integrate the aforementioned concepts to establish a connection between the geometric nature of gravity and the discrete properties of quantum mechanics through an algebraic framework, utilizing the unique perspectives offered by PBHs and the two-dimensional self-gravitating photon BEC. This approach is unique in its synthesis of astrophysical phenomena with aid from quantum statistical mechanics, providing a fertile ground to investigate the quantum aspects of gravity.

Our study aims to illustrate the significance of PBHs in the formation and evolution of the observable universe. It may act as the key factor to solve major questions and paradoxes relevant to Planck-scale physics, such as Big Bang cosmology, the physical background of dark matter and dark energy, baryogenesis, the formation mechanism of SMBHs, early clumpiness of the universe, large-scale structures, such as super-voids and superclusters, and two-stage accelerated expansion of the Universe; the horizon problem, cosmological constant problem, and the information-loss paradox that has arisen in observational cosmology and astrophysics.

By analyzing these specific problems we want to develop a common understanding of how algebraic techniques might shed light on the fundamental properties of spacetime and its quantum features. This method allows us to address theoretical issues and set a framework for future experimental validation.

This manuscript is structured as follows: In section II, we derived the algebra (namely Einstein's algebra) utilizing the membrane formalism, establishing a correspondence between the PBHs and the two-dimensional condensate photons. In section III, we examine the essential foundational principles of physics at the Planck scale. Starting from Section IV, we will utilize our newly developed formalism to explore a diverse array of phenomena, including black hole thermodynamics, luminosity, lifetime, the hot Big Bang, accelerated expansion, and baryogenesis, testing the success of our method to solve some previously unsolved problems of cosmology.  In Section IV we shall elucidate the interconnections between energy, temperature, and entropy of black holes when they approach the Planck scale limit. The discussion of PBHs' lifetime and luminosity, in conjunction with the information loss paradox, is presented in Section V. Subsequently, we outlined the process of free light accretion via quantum absorption, as well as the origins of disks and jets in Section VI. In section VII, we analyze the conservation law of quantum information and the energies associated with gravitational radiation from colliding and merging of black holes, which serve as a driving force behind the accelerated expansion of the universe. In section VIII, we examine the intricate details of Planck-scale phenomena, the dynamics of the hot Big Bang, and the initial phase of the universe's accelerated expansion. In section IX, we derived the connection between the Planckian Era and the Baryogenesis Epoch, addressing the issue of supermassive black hole formation and the primordial clumping of the universe. In section X, we analyze baryogenesis via the mechanism of primordial black hole production. In section XI, we established the methodology for the formation of contemporary dark matter and the large-scale structure of the universe. The final section, XII, serves as the conclusion of this study.

\section{The sketches of ``Einstein's algebra"}

We begin this section with Albert Einstein's final work in 1956 \cite{Einstein}, where he articulated a concluding statement that reflects his core methodological principle: "... a finite system of finite energy can be completely described by a finite set of quantum numbers. This does not seem to align with a continuum theory and must lead to an attempt to find a purely algebraic theory for describing reality. However, nobody knows how to establish the foundation for such a theory".

Inspired by Einstein's vision, we aim to develop a description of physical reality rooted purely in algebra and try to bridge the seemingly insurmountable gap between gravity and quantum mechanics.

As it was mentioned at the beginning of our Introduction, only the gravitating condensed light with its practically limitless range of Compton wavelengths from Planck-scale to tens of diameters of Solar System can serve as the "building material" for black holes. Its quantum nature must be inevitably reflected in the quantized characteristics of black holes. But what is their geometrical structure? What is the constructive design of such unique "buildings" of the observable Universe?

We find a leading answer in the well-established numerical relativity from the well-proven method of the "simple excision of a black hole in 3+1 numerical relativity" \cite{Alcubierre}, where black hole is presented as a very thin layer of gravitating matter, the internal spacetime of which is completely excluded from the calculation.

In this study, we are dealing with a extremely thin "membrane" model of black holes. It is not difficult to imagine what would happen if we replace abstract matter with the experimentally discovered 2D quantum photon condensate predicted by the same A. Einstein a hundred years ago \cite{Klaers}. 

Historically, in 1985, Price and Thorne introduced the "membrane formalism" for black holes \cite{Price}, which reframes the black hole's horizon as a "stretched horizon"—a 2D membrane within 3D space that responds dynamically to external forces. This membrane, building on Damour and Znajek's work \cite{Znajek, Damour, Damour1}, acts like a viscous fluid with electrical charge and conductivity, as well as finite entropy and temperature, though it cannot conduct heat. The membrane’s interactions with the universe are governed by familiar fluid dynamics and thermodynamic principles, such as the Navier-Stokes equation, Ohm's law, and the laws of thermodynamics. These concepts provide astrophysicists with intuitive tools for understanding black hole behavior in complex environments. In this context, we discuss the following two works. The discussion covers two works focused on early universe phenomena. In Grasso et al. \cite{Grasso}, the authors explore the possibility that galactic and cluster magnetic fields have a primordial origin. They suggest that searching for imprints of these magnetic fields in the temperature and polarization anisotropies of the cosmic microwave background radiation (CMBR) is the best approach for testing this hypothesis. 
Another work done by Bruce et al. \cite{Bruce} proposes that dark energy could originate from dark matter condensation at a low redshift. They analyze constraints from CMBR, large-scale structures, supernovae, and radio galaxies using a four-parameter likelihood analysis, factoring in the Hubble constant and three parameters of the condensate field.

The above-mentioned concepts, especially the concept of "membrane paradigm", may also apply to 2D photon Bose-Einstein condensates (BEC) with rest energy (arise due to condensation) in various phenomena \cite{Muller, Klaers,  Klaers1, Jonkers}. The idea of a 2D photon BEC with rest energy and rest mass can be linked to the early universe, particularly concerning PBHs, analog gravity, and related topics \cite{Freitas, Vocke, Liao, Liu}.

The model of PBHs we present in this work is framed as 2D spherical photon condensates contained within their own gravitational fields \cite{Borsevici}. Such a model naturally leads to a geometrical equation that directly connects the Compton wavelength, $\lambda_{sm}$, of condensed light (quantum theory) with the geodesic length (general relativity) \cite{Borsevici}:
\ben
\l_{sm}=2\pi R_{s},
\label{1}
\een
where $R_s$ is Schwarzschild radius. To paraphrase John Wheeler's famous quote, we can say that "The Spacetime tells to condensed Light Quanta how to curve, the condensed Light Quanta tells to Spacetime how to quantize". The expression for Schwarzschild radius in term of mass ($M_s$) of the PBH, may be expressed as 
\ben
R_s=\frac{2 G M_s}{c^2}
\label{2}
\een
In a 2D photon BEC, photons can exhibit behavior analogous to massive particles due to the effective mass they acquire within the system, even though photons are intrinsically massless in free space. This effective mass arises from the interaction of photons with the medium and the trapping potential used to create the condensate. Relating this effective mass($M_{sm}$) to its Compton wavelength ($\l_{sm}$) we can write,
\ben
\l_{sm}=&\frac{2\pi\hbar}{M_{sm}c}
\label{3}
\een
Using the above relationship Eqs. (\ref{1}),(\ref{2}) and (\ref{3}), we promptly determine the effective mass and rest energy of each condensed photon in terms of Compton wavelength ($\lambda_{sm}$) and Schwarzschild radius($R_s$):
\ben
M_{sm} = \frac{2\pi\hbar}{\l_{sm}c} = \frac{\hbar}{R_s c}\n \\ 
E_{sm} = M_{sm}c^2 = \frac{\hbar c}{R_s}
\label{4}
\een
expressing the above quantities in terms of condensate photon number ($N_s$):
\ben
N_s=\frac{M_s}{M_{sm}}=2\frac{M_{s}^{2}G}{\hbar c}=2\frac{M_s^2}{M_p^2}=2\frac{E_s^2}{E_p^2}=\frac{R_s^2}{2l_p^2},~\text{etc.},
\label{5}
\een
expressed in terms of the Planck mass $M_p~(=\sqrt{\frac{\hbar c}{G}})$, Planck energy $E_p$, and Planck length $l_p~(=\sqrt{\frac{\hbar G}{c^{3}}})$. In the above Eq.(\ref{5}), $M_s$ and $E_s$ stands for total rest mass and the rest energy of the condensate photon respectively. These fundamental and consistent relationships may enable us to explore the unknown facets of Planck-scale physics. Considering that the quantity $N_s$ of condensed photons cannot be less than two to form a bound state, we can readily derive a significant expression for $N_s$, utilizing the natural quantum number $n_s$ as
\ben
N_s=2n_s,~n_s=1,2,3,...
\label{6}
\een
Relying solely on established principles of physics, one can readily determine the quantized characteristics of PBHs - such as mass $M_s$, rest energy $E_s$, size $R_s$, area $A_s$, entropy $S_s$, temperature $T_s$, quantum information quantity $I_s$, longevity $\tau_s$, luminosity $L_s$, among others—as well as the fundamental principles governing their formation, evolution, and eventual decay. It is important to note that all these quantized features of PBHs depend exclusively on Planck units and are influenced by various powers of natural quantum numbers \cite{Borsevici}. From the previously established quantum relations we get,

\ben
M_s=M_p\s{n_s};~E_s=E_p\s{n_s};~R_s=2l_p\s{n_s};~\n \\ A_s=4\pi R_s^2=16\pi l_p^2n_s;~S_s=2k_Bn_s;~
T_s=\frac{T_p}{4\s{n_s}};~\n\\ I_s=N_s=2n_s(bit);~\tau_s=\frac{4\pi t_p n_s^{\frac{3}{2}}}{3};~L_s=\frac{L_p}{4\pi n_s},\nonumber\\ 
\label{7}
\een

where $t_p$, $L_p$, and $T_p$ represent the Planckian units of the lifetime, luminosity, and temperature, respectively, and $k_B$ denotes the Boltzmann constant. We can easily find expressions for the above parameters by putting $n_s=1$ for a minimal Planck black hole.\\

To facilitate a direct comparison with the foundational findings of Bekenstein \cite{Bekenstein1} and Hawking \cite{Hawking} for the black hole entropy, we express $n_s$ in terms of the area $A_s$ of the black hole using Eq. (\ref{7}). By substituting the expression for the Planck length $l_p$ into the formulation, we obtain a modified expression that illustrates the connection between quantum states and the geometric characteristics of the black hole as: 
\ben
&&S_s=\frac{k_B c^3}{8\pi\hbar G}A_s=\frac{S_B}{ln2}=\frac{S_H}{2\pi}\n \\
&&T_s=\frac{\hbar c^3}{4k_B GM_s}=2\pi T_H
\label{8}
\een
where we define $S_{B}~(=\frac{k_{B}c^{3}ln2}{8\pi G\hbar}A)$ to represent Bekenstein entropy \cite{Bekenstein1}, $S_{H}~(=\frac{k_{B}c^{3}}{4G\hbar}A)$ to signify Hawking entropy \cite{Hawking}, and $T_{H}~(=\frac{\hbar c^{3}}{8\pi G M k_{B}})$ to denote Hawking temperature. Note that historically Jacob Bekenstein was the first physicist who combined all four universal constants, $G,c,\hbar,k_B$ related to quantum mechanics and gravity in one consistent formula for the black hole entropy.

Looking at the result (\ref{8}) we can state that our findings closely matched to those that Bekenstein and Hawking predicted, which indicates that we are moving in the right direction. However, our findings were derived not from intuitive assumptions, but through clear and exact physical principles of physics.

It should be noted that the quantization approach outlined for Schwarzschild black holes can be naturally extended to Kerr black holes.

The size of the equatorial circumference horizon of a Kerr black hole (the maximal geodesic length) is equal to \cite{Bekenstein4} %,pse27mar2019
\ben
&&K (equatorial~circumference) = \frac{4 \pi G M_k} {c^2}.
\label{9a}
\een
where $M_k$ is Kerr black hole mass, $M_s$ is the mass of the corresponding Schwarzschild black hole taken as the irreducible one, $J$ is angular momentum.  Note that when $J=0$ a Kerr black hole becomes the Schwarzschild one, with its proper form of the geodesic length:
\ben
&&K (equatorial~circumference, J=0) = 2 \pi R_s.
\label{9b}
\een

Putting the natural geometrical equation
\ben
&&\lambda_{km} = \frac {4 \pi G M_k} {c^2}
\label{9c}
\een
and Compton wavelength definition 
\ben
&&\lambda_{km} = \frac {2 \pi \hbar} {M_{km} c}
\label{9d}
\een
we find
\ben
&&n_k = \frac {M_k^2} {M_p^2}
\label{9e}
\een
and other important formulas such as \ben
&&M_k = M_p  \sqrt { n_k },~~ I_k = N_k = 2 n_k,~~ \n \\ && S_k = 2 k_B n_k,~~ T_k = \frac {T_p} { 4 \sqrt { n_k } } etc.
\label{9f}
\een

Starting from well known Kerr black hole formulas \cite{Bekenstein4}:

1) Fundamental relation between $M_k$, $M_s$  and $J$
\ben
&&M_k = \sqrt {M_s^2 + \frac {J^2 c^2} {4M_s^2 G^2}},
\label{9g}
\een

2) The fundamental condition for Kerr black hole outer horizon existence:
\ben
&&J \leq \frac { G M_k^2 } { c },
\label{9h}
\een
we can quantize these relations by the natural substitution:
\ben
&&M_k = M_p \sqrt{ n_k },~~ M_s = M_p \sqrt{ n_s },
\label{9i}
\een
and transform (\ref{9g}) and (\ref{9h}) as follows:
\ben
&&n_k = n_s + \frac { J^2 c^2 } { 4 M_p^4 G^2 n_s },
\label{9j}
\een
\ben
&&J \leq \hbar n_k.
\label{9k}
\een

Note that the right side of this fundamental inequality gives a very interesting physical meaning because we must take into account that the spin of condensed photons is equal to the reduced Planck constant and that the Kerr quantum number is equal to half of the number of condensed photons.

The above results concerning quantized Kerr black holes are of undoubted interest; however, we will not use them further, since the mechanisms of formation of angular momentum, $J$, manifest themselves only in the case of stellar mass and supermassive black holes. 

\section{Planck equivalences and corresponding relations}

It is well-recognized that a hallmark of any well-established theory lies in its natural derivation of fundamental constants, equivalences, and relationships. In our work, we independently developed a significant system of Planck units, necessitating a thorough examination of the physical implications and extensive consequences that arise from it. For instance, this approach reveals a series of astonishing formulas and equivalences demonstrating the system's inherent depth and coherence. The formulas and equivalences are:
\ben
&\text{Planck force:}~F_p=G\frac{M_pM_p}{l_p^2} ~(Newton)=M_p a_p;&\nonumber
\een
\ben
&\text{Planck work:}~W_p=F_pl_p=\frac{\hbar}{c}a_p=E_p;&\nonumber
\een
\ben
\text{Planck energy:}~ E_p &&=M_p c^2~(Einstein)\nonumber \\ &&=\hbar \omega_p~(Planck),~\omega_p=\frac{1}{t_p}; \nonumber
\een
\ben
\text{Planck gravity:}~ && g_p=G\frac{M_p}{l_p^2}=\frac{F_p}{M_p}=a_p~\nonumber \\ &&({\it Einstein~equivalence~principle});\nonumber
\een
\ben
&\text{Planck Momentum:}~P_p=\frac{E_p}{c}=M_pc=\frac{\hbar}{l_p}~(de~Broglie);&\nonumber
\een
\ben
&P_pl_p=E_pt_p=\hbar~({\it Heisenberg~uncertainty~principle});&\nonumber
\een
\ben
\text{Planck Temperature:}~T_p&& =\frac{E_p}{k_B}=\frac{\hbar}{ck_B}a_p \nonumber \\ &&  ~({\it Unruh~relation}), etc., \nonumber \\
\label{9}
\een
where $a_p$ denotes Planck acceleration and $\omega_p$ signifies Planck cyclic frequency, $g_p$ is Planck gravity, and so on. The above relations can be utilized in the subsequent discussion, particularly in the context of the initial state of the observable universe and the processes related to the Big Bang. 

\section{Quantum emission of PBH: Energy and Entropy Radiation}

The description of a black hole closely aligns with the renowned ``membrane formalism" proposed by Price and Thorne \cite{Price}, which reconceptualizes the dynamics of the black hole horizon as a two-dimensional ``stretched horizon" that operates as a physical membrane within three-dimensional space. In our study, we identify two significant differences from that perspective, which warrant careful elucidation \cite{Borsevici}.

1. We introduced the concept of ``membrane stretched horizon" as a two-dimensional photonic condensate.

2. Secondly, in our framework, the gravitational squeezing force, expressed as:
\ben
&&F_s~(gravitational~squeezing ~force)\n \\ &&=\frac{dE_s}{dR_s}= \frac{d}{dR_{s}}(\frac{M_{p}c^{2}R_{s}}{2l_{p}}) \n \\ && =\frac{c^4}{2G}=\frac{F_p}{2}~~
\label{10}
\een
is precisely balanced by analogous repulsive quantum-mechanical counterpart:
\ben
&&F_s~(quantum~mechanical~force)\n \\ &&  =N_s\frac{dE_{sm}}{dR_s} =\frac{R_s^2}{2l_p^2}\frac{d(\frac{\hbar c}{R_s})}{dR_s}=-\frac{F_p}{2}.
\label{11}
\een
This has profound implications for forming a {\it quasi-stable quantum gravitational} state that prevents singularities. An equilibrium between gravitational compression and quantum repulsion, analogous to the effect we found in white dwarf stars where electron degeneracy pressure holds the star against the gravitational collapse, can establish conditions that preserve quantum information at the horizon, consistent with concepts from holography and the AdS/CFT correspondence \cite{Hooft, Maldacena, Susskind, Witten}. The Holographic principle, proposed by Gerard 't Hooft and further developed by Leonard Susskind, proposes that all information contained within a volume can be encoded on its boundary, analogous to a three-dimensional hologram represented on a two-dimensional surface. The AdS/CFT correspondence, introduced by Juan Maldacena in 1999, exemplifies this principle's application. This work helps us to establish a link between quantum field theory and gravity, which makes it easier to study quantum gravity, black holes, and condensed matter physics.

However, this remarkable equilibrium is periodically interrupted by spontaneous quantum transitions, as described by the work of Bekenstein and Mukhanov \cite{Mukhanov2}. This periodic interruption has a profound implication in the context of the structure formation of the universe. According to our understanding, quantum fluctuations are critical in establishing the initial density variations in the early universe that eventually grow into galaxies and large-scale structures. These fluctuations create a ``spontaneous transition" in the nascent universe by introducing new phases or areas with slightly differing energy densities. A ``spontaneous quantum transition" could occur when the universe's energy density reaches a critical threshold, causing it to ``jump" to a new state. This could be analogous to vacuum decay or phase transitions in early cosmology, in which spacetime ``selects" regions with lower energy, resulting in the expansion or reconfiguration of spacetime domains. These transitions occur from an initial
$n_s$ energy level to a lower $n_{s}-1$ state, resembling electron transitions in the hydrogen atom. The conservation of momentum drives such transitions, which involve the emission of two particles ($\gamma_s$) accompanied by gravitational radiation ($\Gamma_s$), in quadrupolar and octupolar wave modes. The fundamental framework governing these transitions is outlined as follows \cite{Borsevici}:

\ben
BH_s\to BH_{s-1}+2\gamma_s+\Gamma_s
\label{12}
\een
with
\ben
E(BH_s)=E_p\s{n_s};~E(BH_{s-1})=E_p\s{n_s-1}.
\label{13}
\een

So that we can write
\ben
\Delta E_{s,s-1}~(rad)=E_p\s{n_s}-E_p\s{n_s-1}
\label{14}
\een
which can also be written as
\ben
&\D E_{s,s-1}~(rad)=\D E_{s,s-1}~qu(rad)+\D E_{s,s-1}~gr(rad)\nonumber\\
\label{15}
\een
where $qu(rad)$ and $gr(rad)$ represents quantum radiation and gravitational radiation respectively. Also, note that throughout the article, ($qu$), ($gr$), and ($rad$) represent ``quantum", ``gravitational", and ``radiation" respectively. We calculate the thermal quantum radiation using Eq. (\ref{7}) as:
\ben
&\D E_{s,s-1}~qu(rad)=2\times \frac{E_p}{4\s{n_s}}~(two-particle~energy)\nonumber\\
\label{16}
\een
and from Eq. (\ref{15}), we get the gravitational radiation as:

\ben
&\D E_{s,s-1}~gr(rad)=\D E_{s,s-1}~(rad)-\D E_{s,s-1}~qu(rad)\n \\&=E_{p}\sqrt{n_{s}}-E_{p}\sqrt{n_{s}-1}-\frac{E_{p}}{2\sqrt{n_{s}}}\n \\&=\frac{E_{p}}{2\sqrt{n_{s}}}\big(\frac{1}{4n_s}+\frac{1}{8n_s^2}+...\big)
\label{17}
\een

This transition resembles those observed in a hydrogen atom, occurring with the emission of two particles to conserve momentum and resulting in gravitational radiation. Gravitational waves arise from the strong gravitational fields of black holes, leading to the distortion of the surrounding spacetime. The most intense gravitational radiation generated in this process exhibits quadrupolar and octopolar features. The coefficients in Eq. ({\ref{17}}), $\frac{1}{4n_s}$ and $\frac{1}{8n_s^2}$, correspond to the quadrupolar and octupolar modes, respectively. From the Clausius relation, we obtain the radiated quantum and gravitational entropy:
\ben
&\D S_{s,s-1}=\frac{\D E_{s,s-1}~(rad)}{T_s~(rad)}=\D S_{s,s-1}~qu(rad)+\D S_{s,s-1}~gr(rad)\nonumber\\
\label{18}
\een
with

\ben
&\D S_{s,s-1}~qu(rad)=2k_B;\n \\ &\D S_{s,s-1}~gr(rad)=2k_B~\big(\frac{1}{4n_s}+\frac{1}{8n_s^2}+...\big),
\label{19}
\een

where $\Delta E_{s,s-1}~(rad)$ - all emitted energy; $\Delta E_{s,s-1}~qu(rad)$ - the energy of two emitted
particles; $\Delta E_{s,s-1}~gr(rad)$ - the energy of gravitational radiation; $T_s~(rad) = T_s$; $\D S_{s,s-1}~(rad)$ - the emitted entropy, that increases the total entropy of the universe; $\D S_{s,s-1}~qu(rad)$ - two
particle radiation entropy, $\D S_{s,s-1}~gr(rad)$ - gravitational radiation entropy. We are typically dealing with the emission of two photons, however, there is one very significant exception to this rule when it comes to baryogenesis (see below).\\

In order to clarify the profound physical implications of the results obtained, it is essential to ``dress" the ``naked" mathematical expression $\D E_{s,s-1}(rad)$ in explicit physical terms.\\
The careful testimony towards Planck equivalences and correlations is helpful to us in the following ways:
\begin{enumerate}
    \item [1.] Considering the straightforward physical expression of the acceleration $a_s$, linked to the dynamical process of a black hole's evaporation using Eq.(\ref{10}):
    \ben
    a_s=\frac{F_s}{M_s}=\frac{\frac{F_p}{2}}{M_p\s{n_s}}=\frac{a_p}{2\s{n_s}},
    \label{20}
    \een
    one can easily find the very interesting physical form that links emitted energy and acceleration, as mentioned above with the Planck equivalence $E_p=\frac{\hbar}{c}a_p$:
    \ben
    &&\D E_{s,s-1}(rad)=\frac{\hbar}{c}a_s+\frac{\hbar}{c}a_s\big(\frac{1}{4n_s}+\frac{1}{8n_s^2}+...\big)
    \label{21}
    \een
    where the first term expresses the energy of two emitted particles, the second term is related to the energy of gravitational radiation, expressed by quadrupolar $\frac{1}{4n_s}$ and octupolar $\frac{1}{8n_s^2}$ terms.
\item[2.] Using the previously stated equivalence $T_p=\frac{a_p}{ck_B}\hbar$, we can derive the remarkable expression that relates the thermal temperature and acceleration as:

\ben
&&T_s(rad)=\frac{E_p}{4k_B\s{n_s}}=\frac{\hbar a_p}{4ck_B\s{n_s}}=\frac{\hbar}{2ck_B}a_s=T_s.
\label{22}
\een

This outcome indicates that the temperature of the quantum state is dependent upon acceleration, complementing Unruh's finding \cite{Unruh} that an accelerated observer in a vacuum perceives it as a warm bath of particles. For a black hole, the term $\sqrt{n_s}$ signifies that the actual temperature varies with the quantum state $n_s$, implying a quantized nature to this thermal phenomenon. This connection to Unruh’s theorem underscores the profound interconnection of quantum mechanics, thermodynamics, and relativity, illustrating how the black holes accelerated evaporation influences the observed properties of the vacuum.
\end{enumerate}

It should be noted that the Unruh theorem \cite{Unruh} played a pivotal role in theoretical physics by enabling T. Jacobson \cite{Jacobson} to derive Einstein’s field equations using thermodynamic principles. Building on this, Jacobson's profound idea, often summarized as ``black hole entropy without black holes"—led to the application of the Clausius relation in this context, connecting black hole thermodynamics with the broader framework of spacetime geometry. The values of gravitational radiation energy and radiated gravitational entropy are substantial only in the Planck scale range, namely during the Planckian era where
quantum numbers commence with the initial natural numbers. However, it is these initial numbers that will unlock the gateway to the beginning of time. This has a significant implication in quantum cosmology according to which the universe’s birth involved quantum gravitational phenomena. The quantization of time, space, and gravitational waves at the planck epoch may imply that classical spacetime evolved from a fundamentally discrete quantum structure. Indeed this idea provides us a way to uncover the quantum structure underlying the universe's beginning and to gain a deeper understanding of time itself.

Note that, following the previous results, the basic relation (\ref{15}) can be easily rewritten in the following forms:
\ben
&&\D E_{s,s-1}(rad) = T_s \D S_{s,s-1};
\n \\ &&
\D E_{s,s-1}(rad) = F_s \D R_{s,s-1};
\n \\ &&
\D E_{s,s-1}(rad) = \frac{F_s}{2\pi} \D \lambda _{s,s-1};
\n \\ &&
\D E_{s,s-1}(rad) = p_s V_{s,s-1}.
\label{23}
\een
where $\D S_{s,s-1}$ is referred to (\ref{18}), $\D R_{s,s-1}=2l_p(\sqrt{n_s}-\sqrt{n_{s-1}})$,  $\D \lambda _{s,s-1}=2\pi \D R_{s,s-1}$ is referred to Compton's wavelength of condensed light (\ref{1}), $p_s=\frac{F_s}{A_s}$ is a PBH pressure, $\D V_{s,s-1}=A_s \D R_{s,s-1}$ is a quantum change of it's volume.

All these unexpected equivalences indicate that the interplay between quantum and gravitational PBH radiation can be articulated through a diverse array of fundamental physical theories, including quantum mechanics, thermodynamics, hydrodynamics, and information theory, all of which incorporate a clearly defined gravitational aspect. 

Moreover, from Eqs. (\ref{17}) and (\ref{23}), it is evident that this comprehensive physical description of the most important cosmic processes is governed by special form of the fundamental Heisenberg uncertainty principle:

 \ben
&&\D E_{s,s-1} \D t_s = \frac{E_p}{2\sqrt{n_s}}\big(1+\frac{1}{4n_s}+\frac{1}{8n_s^2}+...\big) 2\pi t_p \sqrt{n_s}\n \\ &&=\frac{h}{2}\big(1+\frac{1}{4n_s}+\frac{1}{8n_s^2}+...\big)>\frac{h}{2}
\label{24}
 \een
where $h$ is a non-reduced Planck constant.

The fundamental form encompasses gravitational components and is a direct consequence of the algebraic methodology imparted by Albert Einstein in his last lines \cite{Einstein}.

\section{PBHs Lifetime and Luminosity: Addressing the Information Loss Paradox}

We now aim to determine the luminosity and lifetime of PBHs, in light of our newly developed formalism. By using the strong conditions of causality and spherical geometry one can readily calculate the transition time of PBHs from one quantum state to the underlying one. It is equal to half the length of the geodesic line divided by the speed of light \cite{Borsevici},
\ben
\D t_s=\frac{\pi R_s}{c}=2\pi t_p\s{n_s}
\label{25}
\een
The aforementioned expression helps us to calculate the black hole lifetime, which can be readily found using the below-mentioned formula as:
\ben
&&\tau_s=\sum_{n=0}^{n_s}2\pi t_p\s{n}\equiv\int_0^{n_s}2\pi t_p\s{n}dn=\frac{4\pi}{3}t_pn_s^{\frac{3}{2}}\n \\&& =\frac{4\pi G^2M_s^3}{3\hbar c^4}=\frac{\tau_s~(Hawking)}{3840}
\label{26}
\een
and Luminosity
\ben
&&L_s=\frac{\D E_{s,s-1}}{\D t_{s,s-1}}=\frac{\frac{E_p}{2\s{n_s}}}{2\pi t_p\s{n_s}}\n \\ && =\frac{L_p}{4\pi n_s}=\frac{\hbar c^6}{4\pi G^2 M_s^2}=3840 L_s~(Hawking),
\label{27}
\een
where the usual lifetime and luminosity of the black hole can be expressed as: $\tau(Hawking)\simeq\frac{5120\pi G^{2}}{\hbar c^{4}}M^{3}$ and $L(Hawking)\simeq \frac{\hbar c^{6}}{15360\pi G^{2}M^{2}}$ \cite{Hawking, Hawking0, LoPresto}.
Thus our model suggests that black holes undergoing quantum transitions differ substantially from Hawking's semi-classical evaporation model. Quantum processes result in more intense emissions than continuous black-body radiation, reducing the black hole's lifetime by a factor of 3840 and increasing its luminosity by the same factor due to discrete energy jumps. Consequently, without external energy input (e.g., through accretion) to offset this intense quantum emission, the black hole undergoes rapid evaporation. This finding challenges Hawking's assumption that black holes radiate as black bodies under the Stefan-Boltzmann law, showing instead that quantum transitions produce significantly higher radiation intensities.

Now we know from Eq.(\ref{16}) that the energy of each emitted particle is:
\ben
E_{\gamma}(rad)=\frac{\D E_{s,s-1}qu(rad)}{2}=\frac{E_p}{4\s{n_s}}.
\label{28}
\een
Using Eq. (\ref{28}), one can easily find that their emitted wavelength is much more than $R_s$ :
\ben
\l_\gamma(rad)=\frac{2\pi\hbar c}{E_\gamma(rad)}=4\pi R_s>R_s.
\label{29}
\een

Because the wavelength of the emitted particles is greater than the Schwarzschild radius, the PBH's strong gravitational field does not prevent the particles from tunneling out. In our scenario, each quantum transition results in the emission of two particles, each containing a discrete amount of energy and information. As a result, each emission causes the PBH to lose two bits of quantum information. However, in conventional black hole evaporation, Hawking radiation is merely thermal and contains no information about the black hole's internal state, resulting in the so-called information loss paradox. In contrast, our model may encode information in the emitted particles, allowing information about the black hole's interior states to escape with each two-particle emission.  As the PBH emits radiation, it loses mass, energy, and information in a quantized manner, resolving the information loss paradox by allowing information to leave the PBH in the form of emitted quantum bits (qubits). This can be easily deduced from the formula (\ref{19}) as follows:
\ben
I_s(rad)=\frac{\sum_{n=1}^{n_s}\Delta S_{s,s-1}qu(
rad)}{k_B}=2n_s=I_s~~
\label{29a}
\een

\section{Quantum Absorption of Free Light: Origins of Cosmic Disks and Jets}

Considering the quantum characteristics of condensed light in PBHs, it is evident that only free light photons with a wavelength not exceeding the geodesic length of the PBHs are subject to accretion by quantum absorption. In other words, PBHs consume only free light or condensed light (such as from other black holes through collisions and mergers). All ordinary matter is melted in accretion disks and expelled in jets, accompanied by swirling magnetic fields through ``funnel" and ``slingshot" effects.

It is particularly noteworthy that this phenomenon precisely reflects the law of total baryon number conservation. A particular impressive manifestation of these effects, behind which this most important conservation law is hidden, are jets whose length is tens of times more than the size of the galaxies that it produced, such as M87 and Porphyrion. As will be shown below, this is precisely what can explain the exceptionally early formation of the SMBHs from a huge reservoir of free light energy radiated by PBHs.

Referring to the previous section we conclude that the wavelength of an absorbed photon cannot exceed the geodesic length; thus we get a condition on wavelength:
\ben
\l_\g~(absorp.)\leq 2\pi R_s=\l_{sm}
\label{30}
\een
which is equivalent to the energy quantum absorption condition:
\ben
E_\g(absorb)\geq \frac{2\pi\h c}{\l_{sm}}=\frac{\h c}{R_s}.
\label{31}
\een
If, during the specified time interval $\D t_s$ (\ref{25}), which corresponds to the duration of the quantum transition from energy level $n_s$ to $n_{s-1}$, the PBH does not absorb the aforementioned free light quantum of energy, it will continue to evaporate. If it receives, it grows in accordance with the law:
\ben
\D R_{s\g }=\frac{2 G}{c^4}E_\g~(absorb)\geq\frac{2G\h}{c^3R_s}=\frac{l_p}{\s{n_s}}
\label{32}
\een
while the absorption area $A_s$ increases by an amount equal to:
\ben
\D A_{s\g}\geq 16\pi l_p^2.
\label{33}
\een
It is noteworthy that analogous results prompted Bekenstein to formulate his renowned equation for black hole entropy \cite{Bekenstein1}.
However, employing Eqs. (\ref{7}), (\ref{28}), and (\ref{31}) in the context of cosmology, particularly regarding dark matter production, yields the most interesting findings:
\ben
E_{\g}(absorb)\geq 2E_{\g}(rad)
\label{34}
\een
that can be expressed in terms of temperature in the following striking way:
\ben
T_{\g}(absorb)\geq 2T_s.
\label{35}
\een
It means when the temperature of the universe ($T_\g$) falls below twice the temperature of a PBH, it commences evaporation. Furthermore, it is essential to comprehend how, in the nascent universe, the demise of one generation of PBHs facilitated the emergence of the subsequent generation \cite{Khlopov, Hawking, Hawking0, Barrow}. We must adhere to these paramount cosmological criteria in our explanation of the formation of contemporary dark matter, which comprises black holes of asteroid-like mass \cite{Tinyakov}. Another significant consequence of these interactions is that an evaporating PBH can transfer its electromagnetic radiation energy to another PBH only if the mass of the latter is double that of the former \cite{Carr1},
\ben
M_{s_2}(received) \geq 2M_{s_1}(emitted).
\label{36}
\een
These correlations elucidate how, in the primordial phases of the universe's growth, there occurred a cyclic transfer of energy from a dying, more abundant yet lighter generation of PBHs to a new, somewhat less abundant but heavier generation. This is essential for comprehending both the primordial formation of dark matter's structure and the initial ``clumping" of the universe, as these consistent cycles of dark matter generation coincided with the hyper-exponential proliferation of future SMBHs, which served as the centers for the formation of subsequent galaxies.

\section{Conservation of Quantum Information  and Gravitational Radiation from PBH merger}

We find above that two-particle emission from the PBH accompanied by gravitational radiation can potentially resolve the `information loss paradox'. According to the definition provided by John Preskill \cite{Preskill} and Leonard Susskind \cite{Susskind1}, each emitted particle acts as a carrier of one bit of quantum information, consistent with the principles of Shannon entropy \cite{Shannon}. This information is encoded in two possible spin states, $|0>$ and $|1>$, as per the notation established by Paul Dirac \cite{Dirac}. This mechanism helps keep information safe and shows that the unitarity principle in quantum theory is conserved. This means that no information is lost during the evaporation of PBH.

In accordance with the law of information conservation for a binary black hole system ($BH_1$,$BH_2$), which is derived from the fundamental unitarity principle of quantum theory, we can write:
\ben
I_{s1,2}=I_{s1}+I_{s2}=2n_{s1}+2n_{s2}.
\label{37}
\een
From this, we get an interesting formula for the energy of gravitational radiation resulting from the collision and merging of a binary PBH system,
\ben
&&E_{s1,2}(gr(rad))=E_p(\s{n_{s1}}+\s{n_{s2}}-\s{n_{s1}+n_{s2}}).\nonumber\\
\label{38}
\een
This formulation, expressed in terms of quantum numbers, succinctly corroborates Hawking's esteemed area theorem \cite{Hawkings3} concerning the colliding and merging of black holes. During the merger, the resultant black hole experiences a "ringdown" phase, during which it emits gravitational waves while stabilizing into a final state. This process is similar to a struck bell that resonates and then diminishes into quiet, the black hole emits energy as it reaches equilibrium—thus the evocative name ``ringdown."

Examining this phase enables scientists to evaluate predictions of Einstein’s general relativity and to validate ideas such as the area theorem, which posits that the overall surface area of a black hole’s event horizon cannot diminish over time. Recently, a group examining data from LIGO's observations of the ringdown phase presented persuasive evidence \cite{Isi} supporting the validity of Hawking's seminal finding.\\

Now, we obtain the general expression for total gravitational radiation energy for $N$ consecutive collisions and mergers of $N$ PBHs as :
\ben
&&E_{sn}(gr(rad))=E_p\Big(\sum_{i=1}^{N}\s{n_{si}}-\s{\sum_{i=1}^{N}n_{si}}\Big)\n \\ &&=\Big(\sum_{i=1}^{N}M_{si}-\s{\sum_{i=1}^{N}M_{si}^2}\Big)c^2=\Big(\sum_{i=1}^{N}M_{si}-M_{sn}\Big)c^2,
\label{39}
\een
where
\ben
M_{sn}=\s{\sum_{i=1}^{N}M_{si}^2}
\label{40}
\een
is the mass of the resulting black hole. The above Eq. (\ref{39}) with Eq. \ref{40}) is very important in understanding and solving the key problem of cosmology:
\begin{enumerate}
    \item [1.] When the mass of an SMBH at the center of a galaxy greatly exceeds the asteroid-like mass of PBHs situated in the SMBH's ``corona," the dynamics of their mergers yield an interesting outcome. In these instances, nearly all the mass and energy of the tiny PBHs is converted into gravitational wave energy during the merger phase. This process demonstrates a significant transformation in which dark matter is permanently converted into dark energy within the universe. This process serves as a substantial validation of the famous last P. W. Anderson's conjecture, that dark energy is the energy of gravitational radiation \cite{Anderson}. Moreover, as will be shown subsequently, these processes account for the universe's present phase of accelerated expansion.
    \item[2.]  In contrast, Eqs. (\ref{39}) and (\ref{40}) suggest that the formation of an SMBH is unlikely during multiple collisions and mergers of PBHs with nearly equal masses, as most of their total energy would be dissipated as gravitational radiation. It can be inferred that SMBHs primarily originated from the significant light energy associated with radiative processes in the early universe. We will explain below that these critical insights are vital for precisely characterizing and comprehending the substantial role of PBHs and their gravitational radiation in cosmogenesis.
\end{enumerate}

\section{Planck Scale to Expansion: The Quantum Foundations of Cosmology}

The most astonishing aspect of the proposed technique is the logical and inherent emergence of the Planck unit system, devoid of any extraneous or invalid assumptions and hypotheses. Furthermore, from this unforeseen emergence, we discern the authentic principles, laws, and interactions that ruled the observable universe in its primordial condition. This directly facilitates access to the unexplored realm of Planck-scale physics at the inception of time.\\

If we traverse backward through the 'arrow of time', we find spacetime, which is fundamentally three-dimensional and granular; interactions within this framework may lead to photons arranging themselves into a two-dimensional condensate as a lower-energy or stable form. This implies that the 3D spacetime "background" dictates the conditions that allow the 2D condensate to form and persist. These condensed photons having substantial Planck energy and mass densities, are contained within a distinct, non-Euclidean spacetime, characterized by a circumference-to-radius ratio of one. This unique geometry emerges from the substantial Planck gravitational potential, $\phi_p$, which is equivalent to $c^2$, thus confirming the peculiar structure of this domain. Notwithstanding its granular and highly energetic characteristics, this 3D Planck photon condensate establishes an exceptionally dense, spherical arrangement. In a broader perspective, this dense condensate functions as a singular entity with a conventional Euclidean metric, integrating the non-Euclidean characteristics of individual photons into a unified, macroscopic structure.\\

The Planck-Heisenberg uncertainty relation, $E_p t_p=\h $, limits the duration of the observable universe's very unstable starting state to Planck time, $t_p$. Given this tremendous instability, this primordial state soon ``explodes," which is consistent with the predictions of the hot Big Bang theory. This raises several fundamental concerns, like what happens during this explosive event, how long it lasts, and what alterations occur immediately following the Big Bang.\\

In this first stage, we take the universe as a ``granular" spacetime structure that is compact, uniform, and ``granular" \cite{Wheeler1, Hooft1, Rovelli, Bojowald}. Each unit is made up of Planck-scale photons that are packed together into spherical, high-energy packets. Each Planck-condensed photon in this tightly packed, granular spacetime has twelve other photons surrounding it in a very symmetric arrangement. Similar to what we found in FCC structure of atoms/particles, which is the most dense configuration having twelve nearest neighbors in 3D.\\

When the initial state is ``exploding," it is critical to evaluate the possibility of several outcomes for these densely packed photons. For example, what are the chances that a photon with twelve neighbors will break out from its packed configuration? Similarly, how likely is it that groups of twelve surrounding photons will form stable Planck-scale binary systems, either among themselves or with adjacent photons?\\

To quantify these possibilities, one would need to solve a twelfth-degree equation that captures the probability interactions between each photon and its twelve nearest neighbors \cite{Borsevici}:
\ben
\aleph=(1-\aleph)^{12}.
\label{41}
\een
Where $\aleph$ is the probability of converting a condensed Planck photon to a free one and $(1-\N)^{12}$ is the probability for all its twelve surrounding photons to become bound, i.e. to take part at events of the type Eq. (\ref{44}). This equation (\ref{41}) serves to describe the likelihood of certain structural transformations occurring in the aftermath of the explosion, such as the production of photon pairs, chains, or clusters, which might subsequently spread and contribute to the emergent structure of spacetime and matter in the early universe. Solving this twelfth-degree equation would provide a statistical foundation for understanding the complicated dynamics driving the early microstates and their ability to create larger structures or liberate photons in the immediate aftermath of the Big Bang.\\

By solving the above Eq. (\ref{41}) we get the desired probability as: $\aleph=0.1474492855...\approx 0.14745$.\\

Our focus here is on Planck binaries, as their mergers during the Big Bang would give rise to primordial Planck-scale black holes accompanied by Planck-scale gravitational waves. By the energy conservation law, each Planck binary system, comprising two units of Planck mass-energy, would necessarily convert into the mass-energy of a Planck black hole. Additionally, the work involved in the merger process would inevitably result in the emission of Planck-scale gravitational radiation. Mathematically, the entire process is stated as follows:

\begin{enumerate}
\item [1.] With probability $\N$, condensed Planck photons convert to free ones:
\ben
(\g_p) \to \g_p~(free),
\label{42}
\een
the average energy density of free Planck light is
\ben
\Omega_L=\N=0.14745.
\label{43}
\een

\item[2.] Planck binaries ($\g_p,\g_p$) are transformed into Planck black holes ($BH_p$) with probability 1-$\N$, followed by Planck gravitational waves \cite{Borsevici}:
\ben
(\g_p,\g_p) = BH_p + \Gamma_p
\label{44}
\een
where the energy of gravitational radiation ($\Gamma_p$) is caused by Planck work $W_p$.
\ben
E(\G_p)=W_p=F_pl_p=G\frac{M_pM_p}{l_p^2}l_p=E_p;\n\\
E(BH_p)=E(\g_p,\g_p)-E(\G_p)=2E_p-E_p=E_p.
\label{45}
\een

Now we calculate the average Planck dark energy density as
\ben
\Omega_{DE}=\frac{1-\N}{2}\approx0.42628
\label{46}
\een
The average mass/energy density of Planck dark matter is:
\ben
\Omega_{DM}=\frac{1-\N}{2}\approx 0.42628
\label{47}
\een
\end{enumerate}
Take note that current observations indicate that the universe is dominated by dark components, including dark energy and dark matter, which influence its structure and evolution in fundamental ways. Within this framework, it is plausible to consider that the remaining Planck energy density (other than free Planck light energy density) could be statistically divided equally into two distinct contributions: the Planck dark energy density and the Planck dark matter density in its initial stage of formation. This division would suggest that both dark energy and dark matter possess comparable Planck-scale energy densities, potentially offering insights into the underlying nature of these mysterious components that shape the universe. Another aspect is that, by evenly distributing bound energy, we assume that Planck-scale vacuum fluctuations give rise to both dark matter and dark energy states as dual facets of a quantum vacuum, with each contributing equally to the universe's energy content.

So, we found that the Planck energy densities of fundamental free light, dark matter, and dark energy are spread in the following proportion:
\ben
\O_L:\O_{DM}:\O_{DE}\approx 0.14745 : 0.42628 : 0.42628
\label{48}
\een

Note that if we consider the primary 3D Planck photon condensate to be the ``dark" primordial matter, the hot Big Bang process closely resembles the biblical tale of Genesis: ``...and the Creator separated Light from Darkness, and there was the First Day ($\N$)..." \cite{Schroeder}. In the Schroeder \cite{Schroeder} book, he suggests that the description of light's emergence in Genesis aligns with the sudden expansion of light and matter in the Big Bang, paralleling the creation of ``light" and separation of ``darkness" mentioned in the biblical account. Physically, $\aleph$ delineates a universal scale devoid of any extraneous assumptions, reflecting the stability or instability of particle interactions under high-energy conditions around the Planck scale. It elucidates how PBHs and gravitational waves, together with other remnants from this epoch, may contribute to the Lambda-CDM model as large-scale structures arise from Planck-scale physics. $\N$ also plays a significant role in understanding early universe dynamics, marking the likely ``separation" behavior of photons and the proportion of primordial energy that remains unbound—eventually contributing to the observable structure of dark matter and dark energy in our current universe.\\

The so-called ``coincidence problem" may have been resolved in a surprising way at the very beginning of time. The breakdown of extreme homogeneity, which separated the initial ``graininess" of spacetime into Planck-scale binaries and isolated condensed Planck photons, was caused by spontaneous Planck fluctuations. Planck ``granules" acted as ``binary explosives," while Planck fluctuations functioned as ``fuses."\\

It is also possible to compute the duration of the process in two different methods. The wavelength of a Planck-free photon (\ref{29}) is twice that of a Planck condensed photon (\ref{1}), which is equivalent to the Planck length. Naturally, the implementation of such a wavelength necessitates a duration equivalent to $2\pi t_{p}$ (\ref{25}). The second analogous solution arises from the epoch of the formation of Planck black holes. According to the principle of causality, it is equivalent to half the length of its geodesic line divided by the speed of light, specifically $2\pi t_{p}$ (\ref{25}). This apparent alignment of results is likely, not coincidental; rather, it reflects a profound synchronization inherent to all processes at the universe's inception, indicative of the fundamental symmetries and coherence present at the time of the Big Bang.\\

The cosmological constant ($\Lambda$) is a primary parameter in explaining the universe's accelerated expansion. Calculations indicate that $\Lambda$ could theoretically reach a maximum value around $\frac{1}{l_{p}^{2}}\simeq 10^{70} cm^{-2}$, where $l_p$ is the Planck length. This theoretical peak, about $10^{122}$ times greater than today’s observed value, would correspond to early-universe conditions of intense energy density and rapid expansion. This highest value of $\Lambda$ matches a Hubble parameter on the order of $\frac{1}{t_p}$, which is the Planck time. This means that the universe is expanding as quickly as it can under the limits of quantum gravity \cite{Weinberg, Peebles, Padmanabhan, Kiefer}.

The so-called ``vacuum energy" problem needs to be fixed, which said that the cosmological constant was about $10^{120}$ times bigger than what was observed. This is what has been called the ``cosmological constant problem" or ``vacuum catastrophe," which is one of the biggest differences between theory and observation in physics \cite{Carroll, Yoo, Panda}. 

By proposing that $\Lambda$ reaches its maximum value only under early-universe conditions (near the Planck scale), this approach may nullify the enormous factor of $10^{120}$ and indicates that the observed, much smaller value of 
$\Lambda$ reflects a natural evolution from its maximum rather than a fine-tuning issue. This revised perspective potentially resolves the cosmological constant problem by aligning  $\Lambda$ with realistic early-universe scenarios, where its value would have been vastly higher than it is now.\\

Now the question is: what happened during the subsequent ``days of creation"? Subsequent cosmic events start to unfold in full accordance with the principles outlined in the preceding sections. 

The predominant portion of energy from the Planckian free light is absorbed by a fraction, $\N$ (\ref{43}), by the Planck black holes, effectively doubling their mass/energy density. This absorption leads to a reduction in the dark matter mass/energy density to $\O_{DM}=2\N$. Simultaneously, the residual segment of the Planck black holes undergoes disintegration, yielding two photons $\g_{p,0}$ and a gravitational wave $\G_{p,0}$:
\begin{enumerate}

\item [1.] With a probability $\N$ the next generation of black holes is formed by the following scheme:
\ben
\g_p+BH_p\to BH_s~(n_s=4)
\label{49}
\een
Here, to calculate $n_s$, we have to use Eqs. (\ref{13}) and (\ref{45}), and correspondingly 
\ben
M_s=M_p\s{4}=2M_p,
\label{50}
\een
and average (typical) dark matter mass/energy density
\ben
\O_{DM}=2\N\approx 0.29489.
\label{51}
\een

\item[2.] With a probability $\frac{1-\N}{2}-\N=\frac{1-3\N}{2}$ the Planck blackhole decay occures:
\ben
BH_p\to 2\g_{p,0}+\G_{p,0}
\label{52}
\een
with 
\ben
&&2E(\g_{p,0})=2\times \frac{E_p}{4} ,\n\\ &&E(\G_{p,0})=\frac{E_p}{2}
\label{53}
\een
where we have used Eqs. (\ref{16}) and (\ref{52}), and average incremented dark energy:
\ben
\D \O_{DE}=\frac{1-3\N}{2}\times\frac{1}{2}=0.13941,
\label{54}
\een
and the average irreversible dark energy density:
\ben
&&\O_{DE}=\O_{DE}~(precedent ~phase)+\D\O_{DE}\n \\ &&=\frac{1-\N}{2}+\frac{1-3\N}{4}=\frac{3-5\N}{4}\approx 0.56569
\label{55}
\een
and the corresponding average energy $\O_L$, emitted through the Planck $BH_p$ decay to $\g_{p,0}$ photons is
\ben
\O_L=1-\O_{DM}-\O_{DE}=\frac{1-3\N}{4}\approx 0.13941.
\label{56}
\een
However, by the direct calculation, we have the same result as follows: 
\ben
\O_L=\frac{1-3\N}{2}\times 2\times\frac{1}{4}=\frac{1-3\N}{4}\approx 0.13941.
\label{57}
\een

\end{enumerate}
When we compare this second phase (``the second day of creation") with the preceding (Big Bang) one, we find that the dark energy density at Big Bang increased from zero to $0.42628~(=0.56569-0.13941)$, however, during the second part of the Planck epoch, it only grows with the value of $\D\O_{DE}\approx 0.13941$. This implies a significant drop in accelerated expansion in the ``baby universe" (when this increment stops, the universe enters the stage of non-accelerated expansion). We must note that, in line with this process, the density of dark matter decreases from $0.42628$ (\ref{44}) to $0.29489$ (\ref{51}).\\

From this point onward, the gravitational mass of the universe, represented solely by dark matter, will continually decrease, acting as an inexhaustible reservoir of energy for all future generations, leading to the formation of contemporary dark matter, as well as the subsequent emergence of extremely massive and SMBHs and ``ordinary" matter. Let us continue with the examination of the third phase (the ``third day of creation") of the Planckian era. By quantum absorption, the formation of typical black holes $BH_s$ ($n_s$=4) will quickly grow to the standard creation of $BH_s$ ($n_s$=8).\\

Now the question is: What is the basis for this knowledge? According to the preceding phase, the maximum energy reserve for the creation of a typical primordial black hole $BH_s$ ($n_s$=4) via the quantum absorption of free light must be articulated as the total of the prior phase:
\ben
\O_{DM}(max)=\O_L+\O_{DM}=1-\O_{DE}=\frac{1+5\N}{4}\nonumber\\
\label{58}
\een
where we have used Eq. (\ref{55}).
In terms of a typical black hole, this gives the average mass
\ben
M_s=M_p\frac{\O_L+\O_{DM}}{\N}=\frac{1+5\N}{4\N}\approx 2.945498M_p\nonumber\\
\label{59}
\een
Nonetheless, the value 2.945498 does not belong to the series represented by $\s{n_s}$, as it is required to be a conventional quantum number. According to the principle of maximum black hole entropy $S_s$, we determine that the maximum value of $n_s$ that meets the natural requirement is $\s{n_s}<2.945498$.
As a result, we get $n_s$=8, the required number for a typical third-generation primordial black hole.\\

Using previously obtained results, we find the lifetime (using Eq. (\ref{26})):
\ben
\tau_s=\sum_{n=1}^{n_{s}=8} 2\pi t_p\s{n}\approx 32.6\pi t_p,
\label{60}
\een
and emitted free light density (using Eq. (\ref{16})):
\ben
\O_{L}=\N\sum_{n=1}^{n_{s}=8}\frac{1}{2\s{n}}\approx 0.32228.
\label{61}
\een
This is immediately transformed into the $\O_{DM}$ density of the subsequent generation of PBHs via radiative transfer.

The initial findings suggest that taking into account the earlier stages, the total length of the Planckian era is bounded by the value $50\pi t_p \approx 8.46\times10^{-42}$ seconds.

The second result indicates the complete extinction of PBHs that emitted significant light energy, thereby enabling the formation of the next generation. Prior research concerning the principles of quantum absorption indicates that, in the context of radiative transfer between PBH, the mass of the latter must be not less than twice that of the former.

However, what is the source of the final entities? To this point, our analysis has been confined to average, representative values concerning the formation of black holes. In accordance with the foundational concepts of probability theory, it is essential to examine a comparatively small yet impactful quantity of PBHs capable of producing a new generation characterized by mass/energy density,
\ben
\O_{DM} \approx \O_{DL} \approx 0.32228.
\label{62}
\een
As nearly all light energy was absorbed by this generation of PBHs, we arrive at a significant conclusion: by the close of the Planckian era, dark energy approaches its maximum value,
\ben
\O_{DE} \approx 1 - \O_{DM} \approx 0.67771
\label{63}
\een
and its final increment:
\ben
\D\O_{DE}\approx0.67771-0.56569\approx0.11202.
\label{64}
\een

The latter indicates that the conclusion of the Planckian era marks the termination of the initial phase of the universe's accelerated expansion, transitioning to a phase characterized by non-accelerated nevertheless very rapid expansion, as demonstrated below.

Comparing the derived dark energy density value of $0.6777$ with the recent observational data of $0.6847$ from the Planck Collaboration (2020) \cite{Aghanim} suggests that the universe has experienced accelerated expansion over the past five to six billion years due to this increment in dark energy density. This change is given by
\ben
&&\D\O_{DE}=\D\O_{\Lambda}\geq 0.6847-0.6777\n\\ &&=0.0073~(or~greater)
\label{65}
\een
and varies according to differences observed in recent results from various cosmological measurement methods \cite{Aghanim, Riess}. It is also noteworthy that primary gravitational waves (radiation), ($\G_p$ and $\G_{p,0}$), with energies 
$E_p$ and $\frac{E_p}{2}$ respectively, account for approximately $97\%$ of the total dark energy density in the present-day universe, aligning with the findings from recent observational studies \cite{Aghanim}.

It should be noted that, during the Planck era, the thermodynamics of the universe was dictated by the principles of black hole thermodynamics. Contrary to the expectations of a conventional cosmological model, the universe's primordial regions were not causally isolated; instead, it exhibited characteristics akin to black holes, that facilitate thermal equilibrium over all areas. This relationship between black hole thermodynamics and the early cosmos potentially eliminates the necessity for special explanations on how distant regions could possess identical temperatures and characteristics. Thus it may remove the so-called 'horizon problem'.\\

\textbf{Corollarium}. This is an exceptionally important section of our research, which, based on the unified (quantum-mechanical and gravitational) algebraic description of reality bequeathed by A. Einstein, allows us not only to penetrate into unattainable Beginning of Time, but also to understand the great importance of "quantum cosmology" and the accelerated evolution of the Universe. Let us list some important lessons of this research.

1. Moving backwards along the "arrow of time" we find our observable Universe as a "granular" spherical lump sized as an atom of gold, consisting of Planck 3D photon condensate, in a state with extremely high energy and mass density, temperature and pressure, as well extremely low entropy and lifetime, all expressed in Planck units.

2. According to Planck-Heisenberg law, the greatest explosion was occured. As a result of which, during the time 2$\pi$tp, the "grainy" space-time became continuous, Planck free photons, Planck black holes and Planck gravitational waves were formed. It is this moment of the Beginning of Time must be called "hot Big Bang", in which Planck binaries played the role of "explosives", while the Planck fluctuations act as "fuse". The Hubble parameter was close to the limit value 1/tp, Lambda parameter was more than 120 decimal orders of magnitude greater than its current value, which potentially resolve the so-called "cosmological constant problem" or "vacuum catastrophe".

3. The careful analysis of the Planck Era, lasting no more than 50$\pi$tp, clearly shows the absense of baryonic matter and that the thermodynamics of this epoch is determined exclusively by the thermodynamics of primordial black holes, which completely resolves so-called "horizon problem". The absence of baryonic matter clearly indicate that primordial black holes represent the Dark Matter and therefore the irreversible energy of gravitational radiation represents "elusive" Dark Energy. Moreover, this clearly explains the roots of so-called "coincedence problem".

4. It becomes clear that each dying generation of PBHs, through radiative transfer, ensures the growth of a new, less numerous generation of heavier ones, and that this process acquires a cyclical and synchronous character.

\section{Formation of Supermassive Black Holes  and the Universe's Primordial Clumping}

This section addresses the challenge of understanding the formation mechanisms of SMBHs and the initial clumping phenomena observed in the early universe.

As established earlier, successive generations of PBHs formed primarily through the radiative transfer of light energy. This process involved energy transfer from one generation, which was significantly more numerous but ``lighter," to a subsequent generation that was far less numerous but ``heavier." This transfer of energy allowed for the growth and evolution of PBHs over time, leading to the formation of SMBHs at the centers of galaxies. These SMBHs play a crucial role in shaping the structure and dynamics of galaxies. This can be articulated straightforwardly and lucidly with significant implications,
\ben
&&\mathcal{N}_{sj}E_p\s{n_{sj}}=\mathcal{N}_{sj-1}E_p\s{n_{sj-1}}\n \\ &&=\O_{DM}E_u=0.32228\times1.5\times10^{62}E_{p},
\label{66}
\een
where $\mathcal{N}_{sj}$ and $\mathcal{N}_{sj-1}$ represent the average numbers of the PBHs in a given and
previous generations respectively, while $n_{sj}$ and $n_{sj-1}$ refer to the typical quantum numbers representing these generations, $E_u$ is the total energy of the observable universe. It is easy to understand that the time gap between generations corresponds to the lifespan of the preceding generations. According to Eq.(\ref{26}) we can write,
\ben
\D t_{sj,sj-1}=\frac{4\pi}{3}t_p(\s{n_{sj-1}})^3.
\label{67}
\een
Hence, it is easy to observe a direct relationship between the $\s{n_s}$ and the time as:
\ben
\s{n_s}\propto t^{\frac{1}{3}}.
\label{68}
\een
Taking into account that the cosmological scale factor $a(t)$ is proportional to $\s{n_s}$, we find that,
\ben
a(t)\propto \s{n_s}\propto t^{\frac{1}{3}},
\label{69}
\een
so that the Hubble parameter is
\ben
H(t)=\frac{\dot{a}}{a}=\frac{dln(t^{\frac{1}{3}})}{dt}=\frac{1}{3t}
\label{70}
\een
and Hubble radius
\ben
R_H(t)=\frac{c}{H(t)}=3ct.
\label{71}
\een
Our selection of the scale factor $a(t)$ as proportional to $\sqrt{n_s}$ is substantiated by the ideas of Bekenstein and Hawking \cite{Hawking, Bekenstein1}, which assert that entropy is intrinsically associated with the surface area of a black hole. In our proposed model, this surface area is proportional to $n_s$. Therefore, it may be inferred that we should consider taking scale factor a to be proportional to $\sqrt{n_s}$. Alternatively, we can explain the reason as following:  if we consider $a(t) \propto \s{n_s}$, this proportionality allows a simplified relation where changes in $a(t)$ directly reflect variations in the number density in the early universe, to simplify the description of the universe's expansion concerning matter or radiation densities \cite{Weinberg1, Dodelson, Kolb}.

Now we can verify our findings by using the fact that the universe's age and observable radius are \cite{Weinberg1, Carroll1}:
$Age_{U}\approx 8.08\times 10^{60}t_p$ and $R_{U}\approx 2.7 \times 10^{61}l_p$

We find the Hubble Radius at the Universe’s age
\ben
R_H(Age_{U})=3c\times8.08\times10^{60}t_p=2.424\times10^{61}l_p\nonumber\\
\label{72}
\een

Two significant insights can be gained from this unexpected consistency of our results: first, that the suggested method effectively addresses the challenging cosmological problems; and second, that it validates the rapid, non-accelerated expansion of the very early universe ($q=-\frac{\ddot{a}a}{\dot{a}}>0$).

The difference of $-10\%$ between the calculated Hubble radius ($R_{H}(Age_{U})$) and the radius ($R_{U}$) of the present universe clearly shows that the latter has grown a lot because of the second phase of faster expansion. Having comprehended the evolution of dark matter throughout generations, we are prepared to address the puzzling issue of the origin and growth of SMBHs.

Immediately after the end of the Planckian era,  an extraordinary event began to unfold: a few exceptionally rare PBHs gained masses far exceeding those of typical black holes. Each death of the previous generation released an immense burst of light energy, with a small fraction absorbed by the rapidly growing accretion layers of these future cosmic giants.

Based on the fact that both the quantum information and the growing PBH's accretion area are proportional to the quantum number, we can write a growth equation for $n_s(t)$ in accordance with Bohr's correspondence principle as:
\ben
\frac{dn_s(t)}{dt}=B_s(t)n_s(t),
\label{73}
\een
where $B_s(t)$ is the accretion (light absorption) function, expressing the dependence on the radiation of each generation of typical black holes evaporation. We find that
\ben
n_s(t)=e^{\int B_s(t) dt}.
\label{74}
\een
As a result, we obtain (using Eq. (\ref{7})) the exponential growth of the mass of a supermassive black hole:
\ben
M_{SMBHs}(t)=M_{p} e^{\frac{1}{2}\int B_s(t) dt }.
\label{75}
\een

It should be noted that the presented exponential time-dependent mass function may explain the early existence of SMBHs and the most powerful quasars detected by modern telescopes. Again, note that the time-dependent mass function can also be found in dynamical spacetimes like those of the Vaidya spacetime \cite{Vaidya, Husain, Manna}, and aligns with models involving time-dependent mass functions for SMBHs \cite{Carr2}. However, we are not discussing pure gravitational spacetime here.

The growth of it's radius is determined in the same way (Eq.(\ref{7})) as:
\ben
R_{SMBHs}(t)=2 l_{p} e^{\frac{1}{2}\int B_s(t) dt }.
\label{76}
\een

The growth of SMBHs occurs through a cyclical mechanism in which successive generations of PBHs generate significant photon emissions that facilitate the growth of subsequent black holes. Fundamental Eqs. (\ref{16}) and (\ref{27}) describe this cyclic mechanism, which results in significant increases in luminosity with each cycle while the SMBH's accretion area expands at an even faster rate, enhancing its gravitational pull. The gravitational bonding promoted by the growth of SMBHs, acting as gravitational attractors, aggregates the clumps of PBHs into complex structures, which signify the early evolution of dark matter in the early universe. The duration (\ref{26}) of each subsequent cycle increases significantly, thereby improving the formation efficiency of early galactic precursors and SMBHs of different sizes. As these cycles end, the universe cools, leading to a significant reduction in the growth of SMBHs and signaling a shift to a less active cosmological phase.

As detailed in the following section, this final growth phase concludes roughly 190,000 years post-Big Bang. For instance, the supermassive black hole TON~618, with a mass of 66 billion solar masses \cite{Zu}, grew to its current size during its last growth cycle at an average rate of $R_{SMBH}(TON~618) / 190,000~years$, comparable to the speed of a car on a road of average quality. It is important to recognize that the accompanying mass growth rate of $6.9\times 10^{35}kg/year$  represents an unusual cosmic event.

The rapid expansion of SMBHs contributes to early cosmic clustering, which aids in the formation of protogalaxies and, subsequently, early galaxies. Observations from the James Webb Space Telescope (JWST) support this scenario, demonstrating that massive and supermassive black holes act as gravitational attractors, facilitating the formation of galaxies and large cosmic structures in the early universe.

\section{Baryogenesis driven by PBH Evolution}

We intend to provide the most compelling evidence for our model, illustrating that PBHs are directly associated with baryogenesis. It is well-known that dark matter surrounds baryonic matter everywhere \cite{Rubin, Ade, Ferreira}. Furthermore, it turns out that regions with higher concentrations of dark matter on large scales, correspondingly exhibit increased concentrations of ordinary matter.

Strictly speaking, the two-particle emission process as represented in Eq. (\ref{12}) should be written in the following general form:
\ben
BH_s\to BH_{s-1}+X_s+\bar{X}_s+\G_s
\label{77}
\een
where $X_s$ and $\bar{X}_s$ represents particle and antiparticle, $\G_s$ represents gravitational wave.
Since photons are their own antiparticles, we can replace this pair with term $2\g_s$ and obtain previously presented Eq. (\ref{12}).

But, when dealing with the production and emission of various particle-antiparticle pairs, the substitution of two $\gamma_s$ particles remains acceptable due to their inevitable annihilation. The only exception is when CP violations occur owing to certain gravitational influences.

The unresolved puzzles surrounding baryon asymmetry, baryogenesis, and leptogenesis persist as fundamental challenges within the realms of cosmology and particle physics \cite{Dine, Canetti}. The established theories of baryogenesis and leptogenesis exhibit shortcomings in two fundamental aspects. First, they fail to sufficiently consider the observed prevalence of baryonic matter, which constitutes roughly $5\%$ of the total mass-energy of the universe \cite{Aghanim}. Second, they fail to fully explain the precise balance between proton and electron numbers, a symmetry that yields net charge neutrality at both microscopic and macroscopic levels across the ``Cosmic Web" \cite{Davidson}.

Instead, we will take a different approach—moving from facts to theory—by looking at the unique productivity of PBHs and their distinct gravitational and thermodynamic effects.

We propose to treat the observed baryon asymmetry as a fundamental postulate in the same way that Einstein accepted the constancy of the speed of light based on empirical evidence. By taking this asymmetry as a given, we can then explore its implications within the broader framework of cosmological and particle-physics phenomena.

Our focus lies in establishing a correlation between this imbalance and the mechanisms underlying particle pair formation by PBHs, utilizing a systematic approach. This framework makes it possible to find strong links between asymmetry in the real world and mechanisms that work in theory. This helps us learn more about how particles form in the early universe and when gravity is very strong.

In the high-energy environment of the early universe, neutrons and protons could freely convert into one another due to the abundant energy available \cite{Mukhanov3, Weinberg2}. However, as the universe expanded and cooled, the energy levels dropped, particularly when positron-electron ($e^+-e^-$) annihilation became significant and neutron decay to protons became more prevalent, as lower available energy could no longer support reverse conversions \cite{Kolb}. In this context, PBHs start to play a decisive role in the production of ordinary matter through decay processes.

The productive decay of PBHs can be represented as follows:
\ben
BH_s\to BH_{s-1}+2n^0+\G_s
\label{78}
\een
where each decay step produces two neutrons ($n^0$)  and an gravitationl energy ($\G_s$), associated with the process. Subsequently, the neutrons decay further:
\ben
2n^0\to 2p^{+}+2e^{-}+...
\label{79}
\een
Here, $n^0$, $p^+$ and $e^-$ represent neutrons, protons, and electrons respectively. Through these sequential interactions, PBHs can contribute to the production of baryonic matter, aiding in the synthesis of fundamental particles necessary for nucleosynthesis and the eventual formation of atomic nuclei \cite{Hawking1, Weinberg, Weinberg1, Weinberg2}. Now, we can easily determine the restriction imposed on the ``productive" quantum numbers of the respective PBHs following Eq. (\ref{28}),
\ben 
\frac{E_p}{4E(c_q)}\leq \s{n_s}\leq \frac{E_p}{4E(n^0)}
\label{80}
\een
where $E(n^0)=0.9396GeV$ is the rest energy of the neutron, $E(c_q)=1.25GeV$ energy of the free ``charm" quark, next in this line after neutron. This constraint implies that PBHs predominantly generate neutrons up to a specific quantum energy level $n_s$. The highest limit, $E(c_q)$, ensures that the PBHs won't have enough energy to produce charm quarks, restricting particle production to neutrons and lighter particles. By limiting $n_s$ to the specified range, the model ensures that the PBH decays predominantly yield baryons, such as protons, rather than other particles that would not enhance baryon asymmetry. The concept assumes that gravitational effects or PBH decay can induce CP violation, essential for generating a matter-antimatter imbalance. PBHs create conditions in which gravitational interactions may inherently result in such violations, eliminating the necessity for alternative CP-violating mechanisms. Here we need the ``charm" quark rest energy in order to determine the quantum number $n_s(c_q)$ at which neutron production stops. Using a simple equation \cite{Borsevici3}
\ben
\frac{[2n^*_s(n^0)-2n_s(c_q)]E(n^0)}{E_p\s{n_s^*(n^0)}}=\frac{\O_b}{\O_{DM}}
\label{81}
\een
(where $\O_b$ is modern baryonic density) we can determine the typical boundary on $n_s^*(n^0)$, which characterises the generation of black holes capable of producing ``ordinary matter" of the observable universe.

The expression on the left side of Eq. (\ref{81}) denotes the productivity of one typical PBH, expressed as the difference between the number of neutrons and the number of annihilated charm quarks it produces (square brackets) multiplied by the rest of the energy of the neutron and divided by the rest energy of typical producing black hole. We equate this to the energy density of baryon $\O_b$=0.0496 divided by the energy density of the producing dark matter $\O_{DM}$=0.32228 (Eq. (\ref{62})). In essence, this equation as a whole expresses a delicate balance between the productive capacity of PBHs in generating baryonic matter and the observed ratio of baryonic to dark matter in the universe. 

Here, we find a quadratic equation for $\s{n_s^*(n^0)}=x$ as:
\ben
x^2-\frac{\O_b}{\O_{DM}}\frac{E_p}{2E(n^0)}x-n_s(c_q)=0
\label{82}
\een

Solving the above Eq. (\ref{82}) by using $\s{n_s(c_q)}=\frac{E_p}{4E(c_q)}$ and the values of $\O_{DM}$, $\O_b$ and $E_p=1.22\times 10^{19}$GeV, we find that $\s{n_s^*(n^0)}=2.9902\times 10^{18}$ corresponding to energy $E^*(n^0)=1.02 GeV$. Here, we disregard the negative root of the quadratic equation, as the quantum number $n_s$ cannot take negative values. Consequently, ordinary (baryonic) matter production from PBHs is feasible within the energy range of $1.02$ to $1.25$ GeV.

According to Eq. (\ref{26}) the formation of baryonic matter takes place over roughly $87$ thousand years, while the complete cycle of PBH evaporation and subsequent extinction lasts approximately $190$ thousand years. During the intermediate period, from $87$ to $190$ thousand years, PBHs may continue decaying and emitting radiation, even though baryonic matter production has largely ceased. This radiation increases the universe's overall radiation density but does not affect the baryon-to-dark matter ratio, which was established during the initial baryonic production phase \cite{Carr3}, in strict conformity with the baryon number conservation law.

To determine the total period during which neutron-to-proton/electron conversion occurs, we need to calculate the duration of proton/electron generation. Based on the above information, we can infer that this conversion process begins when the available energy decreases below the rest mass energy of the charm quark ($1.25$ GeV) and concludes once the available energy reaches the threshold value (i.e., $1.02$ GeV) for baryonic matter formation. For this process, the corresponding quantum numbers are $\s{n_{si}}=\frac{E_p}{4E(c_q)}=2.44\times 10^{18}$ and $\s{n_{sf}}=2.9902\times 10^{18}$ (Eq.(\ref{79})), where $n_{si}$ and $n_{sf}$ stand for the initial and final quantum number for the generation of baryonic matter.

This duration corresponds exactly to half of the cosmic interval from the Big Bang to the Recombination epoch, around $378$ thousand years \cite{Aghanim, Bennett}. This epoch also witnessed the final significant reheating event in the universe. As the universe cooled, ordinary matter settled into a relatively hot plasma state, enabling primary nucleosynthesis, during which protons and neutrons combined to form the first atomic nuclei, signaling the beginning of a phase in which stable elements could emerge.

The most important observation is that the processes involved in the generation of baryonic matter, the evaporation of PBHs, and the reheating phase occurred in localized ``clumps" rather than in a uniform manner. This non-uniformity played a significant role in the subsequent formation of the first galaxies, suggesting that before the phase of ordinary matter creation, the universe likely displayed inhomogeneities. The quantum perturbation \cite{Mukhanov3} could also give rise to these inhomogeneities. Primordial density fluctuations can lead to the formation of gravitational wells that attract matter, thereby increasing particle interactions, promoting the formation of PBHs, and resulting in the production of matter within those regions. Consequently, primordial galaxies emerged in proximity to these dense regions, where PBHs facilitated baryonic synthesis and localized reheating occurred, thus establishing the foundation for the large-scale structure of the Universe.

These results are important for preserving the well-deserved reputation of the standard $\Lambda$CDM cosmological model by freeing it from the arbitrarily introduced so-called "hierarchical" models, that contradict JWST data.

\section{The Evolution of Dark Matter and Its Impact on Cosmic Structure}

The mechanisms of baryogenesis and leptogenesis, along with the continuous radiative transfer of light energy from the last generation of evaporated PBHs to their contemporary equivalents, persist indefinitely.

The formation process ended when the temperature inside the dense clumps dropped below the threshold level of $511~KeV/k_B$, which is the electron rest energy, and then cooled to a lower equilibrium value of $80~ KeV/k_B$ \cite{Carr4}. At this stage, the opacity within the medium was significantly reduced, stopping the growth of PBH, which is posited as the beginnings of dark matter \cite{Green}. Consequently, these PBHs evaporated through Hawking radiation, emitting high-energy X-rays into the surrounding environment.

By applying the Eqs. (\ref{7}), (\ref{12}) and (\ref{16}) that relate the threshold accretion temperature to the mass of a black hole and its associated parameters, we derive the following typical characteristics of contemporary dark matter: 
\ben
&&\s{n_s}=\frac{E_p}{2\times80 KeV}=7.625\times10^{22}~(typical),\n \\&&
M_s=1.659\times10^{18} g~(typical);
\label{83}
\een
\ben
&&\s{n_s}=\frac{E_p}{2\times511 KeV}=1.194\times10^{22}~(minimal)\n\\ && M_s=2.598\times10^{17}g~(minimal),
\label{84}
\een
where we have to use Planck mass $M_p=2.176\times 10^{-5}g$.

The hard X-ray spectrum emitted by PBHs offers valuable insights into the nature of dark matter and the mechanisms potentially driving cosmic acceleration. According to Eqs. (\ref{34}), (\ref{83}), (\ref{84}) this emission spectrum is characterized by a peak intensity at photon energies near $E_{\g}=30~KeV$ and typical intensity of $E_{\g}=40~KeV$, with an upper cutoff observed at around 
$250~KeV$. This exactly matches the unresolved origin of the cosmic hard X-ray background (CXRB) \cite{Bambi, Kahn}, which was discovered before the famous cosmic microwave background (CMB). The spectral properties are important as they may clarify the mass distribution of PBHs and impose constraints on dark matter theories, considering the proposed role of PBHs as candidates for dark matter \cite{Carr, Green}. Furthermore, the X-ray emissions contribute to our understanding of high-energy processes in the early universe, influencing theories on the acceleration of cosmic expansion.

This distinctive hard X-ray spectrum has been observed in high-energy environments across the cosmos, including active galactic centers, galaxy clusters, and superclusters \cite{Bambi, Kahn, Hickox}. The extensive emissions indicate the potential existence of a substantial population of PBHs, supporting the theory that PBHs may represent a considerable fraction of dark matter \cite{Carr, Green2}. PBHs may crash into and merge in the hot, high-energy parts of SMBH coronas because of the strong gravitational fields around them. This can release a lot of X-rays. The resulting intense X-ray radiation, especially in non-jet-related phenomena, aligns with the theory that PBH activity may account for a considerable portion of the unresolved cosmic X-ray background \cite{MacGibbon, Page, Cappelluti}.

Thus, we may conclude that the hard X-ray spectrum produced by PBHs offers essential information regarding the characteristics of dark matter, and the mechanisms responsible for X-ray emission in regions of high density, which also play a role in the generation of dark energy. These interactions enhance our comprehension of PBHs as candidates for dark matter and indicate their potential importance in the universe's accelerating expansion, illustrating the relationship between dark matter and dark energy in cosmic evolution. This transition of dark matter into dark energy may possibly resolve the ``core-cusp problem" \cite{Blok}  which refers to a discrepancy between the predicted and observed density profiles of dark matter in the centers of galaxies, particularly in galaxy clusters and dwarf galaxies.

The JWST has revealed unexpectedly mature galaxies in the early universe, that strongly support the above-presented results, and, which may not be explained by the arguments, presented in \cite{Ferrara}. The fast growth of dark matter halos, which create gravitational wells, would naturally speed up the formation of the universe's structure by directing normal matter into areas with lots of it, where galaxies and stars may come together faster than what was thought before. Consequently, these well-structured galaxies, identified in what was previously considered a primordial and underdeveloped epoch, exemplify the dynamic efficacy of the universe's formative mechanisms.

To the best of our knowledge, we can say that black holes with masses around $10^{18}$~g (Eq.(\ref{80})) have lifetimes much more than the age of the Universe (at about 50 million times more!).

It is clear from the above statement that the average number density of dark matter particles (or PBHs) is about $10^{41}$ times lower than that of normal matter when we look at the current mass parameter $M_s(\sim 10^{18} g)$ along with the known values of $\O_b$ and $\O_{DM}$. This can be calculated by the formula $\frac{\O_{DM} m_p}{\O_b M_s}$, where $m_{p^+}=1.67\times 10^{-24}g$ is the mass of a proton. This aligns perfectly with the cosmic phenomenon of the ``Bullet Cluster" \cite{Markevitch}, where vast clouds of dark matter move through each other without obstruction, demonstrating their minimal interaction. In contrast, intergalactic gas clouds, which possess a notably higher density and are rich in ordinary matter, experience intense electromagnetic interactions that lead to their deceleration and dispersion when they collide. This notable disparity underscores the intriguing properties of dark matter, which interacts exclusively through gravitational forces.

It is important to note that, in the initial epochs of the universe, galaxies were closely clustered, forming a substantial "universal supercluster" filled with dark and baryonic matter. As the universe expanded rapidly, fractures and separations emerged, leading to the formation of the massive cosmic structures we observe today—immense voids, superclusters, great filaments, and walls. The important features arise as natural consequences of this rapid expansion, logically derived from the previously mentioned results.

\section{Conclusion}

Our article examines PBHs as crucial factors in the universe's evolution, specifically the early universe--relating to the Planck scale physics, particularly in relation to quantized gravity and cosmic events. The suggested model links the quantization of black hole characteristics with fundamental physics and cosmology, providing insights into high-energy astrophysics, and cosmic structure formation. This concept achieves the unification of quantum physics and gravity by the groundbreaking discovery of condensed light, namely, the self-gravitating photon Bose-Einstein condensate. This technique is fundamentally based on a geometric understanding that connects the Compton wavelength of the photon condensate to the gravitational radius of PBHs. This connection elucidates a range of quantized attributes-encompassing geometrical, gravitational, thermodynamic, informational, and radiative properties-illuminating the fundamental principles that dictate the formation, development, and ultimate extinction of successive generations of PBHs. 

This advancement unveiled new aspects of Planck-scale and PBHs physics, facilitating the creation of efficient and robust mathematical instruments. These algebraic tools, coined by A. Einstein in his last lines, have facilitated accurate computations and resolutions to various fundamental issues in physics and cosmology, encompassing the Beginning of Time, Big Bang cosmology, the genesis and development of dark matter and dark energy, as well as the mystery of baryogenesis and baryonic asymmetry. Our model also examines the hyper-exponential growth of SMBHs, the birth of primordial galaxies, the evolution of large-scale structures, and the Universe's two-phase accelerated expansion.

Furthermore, we offer a very plausible solutions for multiple persistent issues, problems, paradoxes, and "mysteries" such as cosmological constant problems, informational loss paradox, cosmological principle problems related to the large-scale structure ”excesses”, physical nature and origin of dark matter and dark energy, baryogenesis and baryonic assymetry, the very early growth of SMBHs as the gravitational centers of the very early proto-galaxies, the horizon problem, the core-cusp dilemma, the very early clumping of the Universe and it’s accelerated evolution from the Beginning of Time.

The most notable aspect of our results is their remarkable alignment with the predictions and observations of $\Lambda$CDM cosmology, which has been supported by many observations, including data from the JWST. This agreement is especially significant since our findings were obtained independently, without dependence on the conventional equations and assumptions often linked to $\Lambda$CDM.

The proposed framework provides a systematic method for analyzing quantum-gravitational processes at the Planck scale and addresses specific cosmological phenomena. However, its assumptions and predictions require further experimental and observational confirmation and comparison with other models. The main objective of this work is to provide the principal direction for further theoretical investigations and space exploration.\\

{\bf Acknowledgement:}
The authors extend their gratitude to Prof.~R.~Ruffini, Prof.~G.~Vereshchagin, Prof.~V.~Mukhanov, Prof.~E.~Guendelman, Prof.~S.~Ansoldi, Prof.~A.~Pitteelli, Prof.~A.~Gaina, Prof.~V.~Galamaga, Prof.~P.~Panchadhyayee, I.~Borşevici, and S.~Borşevici for their valuable and insightful discussions, which greatly contributed to the development of this work.
V.B. extends his gratitude to the organizing committee and session chairpersons of the parallel session titled "Inflation: Perturbations, Initial Singularities, and Emergent Universes" at the 17th Marcel Grossmann Meeting in Pescara, Italy, for the opportunity to present his talk at the event. This paper is the comprehensive and thoroughly revised version of the MG17 presentation. G.M. would like to extend thanks to all the undergraduate, postgraduate, and doctoral students who significantly enriched him.\\

{\bf Conflicts of interest:} The authors declare no conflicts of interest.\\

{\bf Data availability:} There is no associated data with this article, and as such, no new data was generated or analyzed in support of this research.\\

{\bf Declaration of competing interest:}
The authors declare that they have no known competing financial interests or personal relationships that could have appeared to influence the work reported in this paper.\\

{\bf Declaration of generative AI in scientific writing:} The authors state that they do not support the use of AI tools to analyze and extract insights from data as part of the study process.\\


\begin{thebibliography}{99}
\bibitem{Green}
A. M. Green, ``PBHs as a dark matter candidate - a brief overview", Nucl. Phys. B, {\bf 1003}, 116494, (2024), \url{https://doi.org/10.1016/j.nuclphysb.2024.116494} 

\bibitem{Choudhury}
S. Choudhury and M. Sami, ``Large fluctuations and PBHs", 	arXiv:2407.17006, (2024) \url{https://doi.org/10.48550/arXiv.2407.17006}

\bibitem{Khlopov}
M. Y. Khlopov and  A. G. Polnarev, ``PBHs as a cosmological test of grand unification", Phys. Lett. B, {\bf 97}, 3, 383-387, (1980), \url{https://doi.org/10.1016/0370-2693(80)90624-3}

\bibitem{carr}
B. J. Carr et al. ‘Observational Evidence for Primordial Black Holes: A Positivist Perspective’. Phys. Rep. , {\bf 1054}, 1–68, (2024) , \url{https://doi.org/10.1016/j.physrep.2023.11.005}

\bibitem{Carr}
B. Carr, and F. Kühnel, ``PBHs as Dark Matter: Recent Developments", Annu. Rev. Nucl. Part. Sci., {\bf 70}, 355-394, (2020), \url{https://doi.org/10.1146/annurev-nucl-050520-125911}

\bibitem{Khlopov1}
M. Khlopov, ‘Primordial Black Holes’, Research in Astronomy and Astrophysics, {\bf 10}, 6 , 495–528, (2010), \url{https://doi.org/10.1088/1674-4527/10/6/001}

\bibitem{Khlopov2}
 M. Khlopov, ‘Primordial Black Hole Messenger of Dark Universe’, Symmetry, {\bf 16}, 11, 1487, (2024), \url{https://doi.org/10.3390/sym16111487}

\bibitem{Belotsky1}
 K. M. Belotsky  et al. ‘Signatures of Primordial Black Hole Dark Matter’. Mod. Phys. Letters A, {\bf 29}, 37, 1440005, (2014) ,\url{https://doi.org/10.1142/S0217732314400057}

\bibitem{Belotsky2}
 K. M. Belotsky et al. ‘Clusters of Primordial Black Holes’. The Eur. Phys. J. C, {\bf 79}, 3, 246, (2019),\url{https://doi.org/10.1140/epjc/s10052-019-6741-4}

\bibitem{Hawking1}
B. J. Carr and S. W. Hawking, ``Black Holes in the Early Universe",  Month. Not. Royal Astro. Soc., {\bf 168}, 2, 399–415, (1974) \url{https://doi.org/10.1093/mnras/168.2.399}



%\bibitem{Euler}
%H. Euler and V Heisenberg,  ``Consequences of Dirac's Theory of Positrons." Z. Phys., {\bf 98}, (11–12), 714-732, (1936) \url{https://doi.org/10.1007/BF01343663}



\bibitem{Muller}
E. E. Müller, ``Note on BEC of photons", Phys. Rev. A {\bf 100}, 053837, (2019), \url{https://doi.org/10.1103/PhysRevA.100.053837}

\bibitem{Ruffini}
R. Ruffini and S. Bonazzola, ``Systems of self-gravitating particles in general relativity and the concept of an equation of state." Phys. Rev., {\bf 187}, 5, 1767-1783, (1969) \url{https://doi.org/10.1103/PhysRev.187.1767}

\bibitem{Klaers}
J. Klaers et al., ``BEC of photons in an optical microcavity", Nature, {\bf 468}, 545–548, (2010) \url{https://doi.org/10.1038/nature09567}

\bibitem{Belgiorno}
 F. Belgiorno et al., ``Analog Hawking effect: BEC and surface waves", Phys. Rev. D {\bf 102}, 105004, (2020), \url{https://doi.org/10.1103/PhysRevD.102.105004}

\bibitem{Maldacena}
J. Maldacena,  ``The Large N Limit of Superconformal Field Theories and Supergravity" Int. Jour. Theor. Phys, {\bf 38}, 4, 1113–1133, (1999), \url{https://doi.org/10.1023/A:1026654312961}



%\bibitem{Wheeler}
%J. A. Wheeler, ``Geons." Phys. Rev., {\bf 97}, 2, 511-536, (1955) \url{https://doi.org/10.1103/PhysRev.97.511}



\bibitem{Schley}
R. Schley et al., ``The Planck distribution of phonons in a Bose-Einstein condensate", 	Phys. Rev. Lett. {\bf 111}, 055301 (2013), \url{https://doi.org/10.1103/PhysRevLett.111.055301}

\bibitem{Bengochea}
G. R. Bengochea, G. Leon and A. Perez, ``Is Planckian discreteness observable in cosmology?", arXiv:2405.12534, (2024), \url{https://doi.org/10.48550/arXiv.2405.12534}

\bibitem{Bekenstein1}
J. D. Bekenstein, ``Black Holes and Entropy", Phys. Rev. D {\bf 7}, 2333, (1973), \url{https://doi.org/10.1103/PhysRevD.7.2333}

\bibitem{Hawking}
S. W. Hawking, ``Particle creation by black holes", Commun. Math. Phys. , {\bf 43}, 3, 199-220, (1975), \url{https://doi.org/10.1007/BF02345020}


\bibitem{Vretenar}
M. Vretenar, C. Toebes and J. Klaers, ``Modified BEC in an optical
 quantum gas", Nature Communications, {\bf 12}, 5749, (2021), \url{https://doi.org/10.1038/s41467-021-26087-0}

\bibitem{Kirton}
P. Kirton et al., ``Introduction to the Dicke model: from equilibrium to nonequilibrium, and vice versa", Adv. Quant. Techno., {\bf 2}, 1, 1800043, (2019), \url{https://doi.org/10.1002/qute.201800043}

\bibitem{Mazur}
 P. O. Mazur and E. Mottola, ``Gravitational Condensate Stars: An Alternative to Black Holes",  Universe, {\bf 9}, 88, (2023), \url{https://doi.org/10.3390/universe9020088}

\bibitem{GW1}
B. P. Abbott et al., {\it LIGO Scientific Collaboration and Virgo Collaboration}, ``Observation of Gravitational Waves from a Binary Black Hole Merger", Phys. Rev. Lett., {\bf 116}, 6, 061102, (2016), \url{https://doi.org/10.1103/PhysRevLett.116.061102}

\bibitem{GW2}
B. P. Abbott et al., ``Multi-messenger Observations of a Binary Neutron Star Merger",  Astrophys. Jour. Lett., {\bf 848}, 2, L12, (2017), \url{https://doi.org/10.3847/2041-8213/aa91c9}

\bibitem{GW3}
B. P. Abbott et al., {\it LIGO Scientific Collaboration and Virgo Collaboration}, ``GWTC-1: A Gravitational-Wave Transient Catalog of Compact Binary Mergers Observed by LIGO and Virgo during the First and Second Observing Runs", Phys. Rev. X {\bf 9}, 031040, (2019), \url{https://doi.org/10.1103/PhysRevX.9.031040}

\bibitem{GW4}
R. Abbott et al., {\it LIGO Scientific Collaboration, Virgo Collaboration, and KAGRA Collaboration}, ``GWTC-3: Compact Binary Coalescences Observed by LIGO and Virgo during the Second Part of the Third Observing Run", Phys. Rev. X, {\bf 13}, 041039, (2023), \url{https://doi.org/10.1103/PhysRevX.13.041039}

\bibitem{Naidu}
R. P. Naidu et al., ``Two Remarkably Luminous Galaxy Candidates at $z \simeq 10–12$ Revealed by JWST", Astrophys. Jour. Lett., {\bf 940}, L14, (2022), \url{https://doi.org/10.3847/2041-8213/ac9b22}

\bibitem{Napolitano}
L. Napolitano et al., ``Seven wonders of Cosmic Dawn: JWST confirms a high abundance of galaxies and AGNs at $z \simeq 9-11$ in the GLASS field", 	arXiv:2410.10967, (2024), \url{https://doi.org/10.48550/arXiv.2410.10967}

\bibitem{Labbe}
I. Labbe et al., ``A population of red candidate massive galaxies ~600 Myr after the Big Bang", Nature, {\bf 616}, 266–269, (2023), \url{https://doi.org/10.1038/s41586-023-05786-2}

\bibitem{Wang}
F. Wang et al., ``A Luminous Quasar at Redshift 7.642", Astrophys. Jour. Lett., {\bf 907}, L1, (2021), \url{https://doi.org/10.3847/2041-8213/abd8c6}

\bibitem{Einstein}
A. Einstein, {\it The Meaning of Relativity (Relativistic Theory of the non-symmetric Field)}, 5th edn. Princeton University Press, Princeton, New Jersey, (1956), \url{chrome-extension://efaidnbmnnnibpcajpcglclefindmkaj/https://lectures.princeton.edu/sites/g/files/toruqf296/files/2020-08/_Albert_Einstein__Brian_Greene__The_meaning_of_rel_BookZZ.org_.pdf}

\bibitem{Price}
R. H. Price and  K. S. Thorne, ``Membrane viewpoint on black holes: Properties and evolution of the stretched horizon", Phys. Rev. D {\bf 33}, 915, (1986), \url{https://doi.org/10.1103/PhysRevD.33.915}

\bibitem{Znajek}
R. L. Znajek, ``The electric and magnetic conductivity of a Kerr hole", Month. Not. R. Astron. Soc., {\bf 185}, 4, 833–840, (1978), \url{https://doi.org/10.1093/mnras/185.4.833}

\bibitem{Damour}
T. Damour, ``Black-hole eddy currents", Phys. Rev. D {\bf 18}, 3598, (1978), \url{https://doi.org/10.1103/PhysRevD.18.3598}

\bibitem{Damour1}
T. Damour, in Proceedings of the Second Marcel Grossman
 Meeting on General Relativity, edited by R. Ruffini (North
Holland, Amsterdam, 1982, p. 587, \url{https://inis.iaea.org/search/searchsinglerecord.aspx?recordsFor=SingleRecord&RN=14722556}

\bibitem{Grasso}
D. Grasso and H. R. Rubinstein, ``Magnetic fields in the early Universe", Phys. Rep., {\bf 348}, 3, 163-266, (2001), \url{https://doi.org/10.1016/S0370-1573(00)00110-1}

\bibitem{Bruce}
A. Bruce et al., ``Condensate cosmology: Dark energy from dark matter", Phys. Rev. D {\bf 68}, 043504, (2003), \url{https://doi.org/10.1103/PhysRevD.68.043504}

\bibitem{Klaers1}
J. Klaers, F. Vewinger and M. Weitz, ``Thermalization of a two-dimensional photonic gas in a ‘white wall’ photon box", Nature Phys., {\bf 6}, 512–515, (2010), \url{https://doi.org/10.1038/nphys1680}

\bibitem{Jonkers}
J. Jonkers, ``High power extreme ultra-violet (EUV) light sources for future lithography", Plasma Sources Sci. Technol. {\bf 15}, S8–S16, (2006), \url{https://doi.org/10.1088/0963-0252/15/2/S02}



%\bibitem{Misner}
%C. W. Misner, K. S. Thorne and J. A. Wheeler, ``Gravitation", Princeton University Press, Oxford, England, (2017)

%\bibitem{Ashtekar}
%A. Ashtekar, and M. Bojowald,   ``Black hole evaporation: A paradigm" 	Class. Quant. Grav., {\bf 22}, 16, 3349-3362, (2005), \url{https://doi.org/10.1088/0264-9381/22/16/014}

%\bibitem{Bekenstein}
%J. D. Bekenstein, ``The quantum mass spectrum of the Kerr black hole" Lett. Nuovo Cim., {\bf 11}, 467-470,  (1974), \url{https://doi.org/10.1007/BF02762768}

%\bibitem{Hawking2}
%S. W. Hawking, ``Gravitationally Collapsed Objects of Very Low Mass", Month. Not. Royal Astro. Soc., {\bf 152}, 1, 75–78, (1971), \url{https://doi.org/10.1093/mnras/152.1.75}



\bibitem{Freitas}
R.C. Freitas et al., ``Primordial Universe with radiation and Bose-Einstein condensate",  Phys. Dark Univ.,  {\bf 25}, 100325, (2019), \url{https://doi.org/10.1016/j.dark.2019.100325}

\bibitem{Vocke}
D. Vocke et al., ``Rotating Black Hole Geometries in a Two-Dimensional Photon Superfluid", Optica, {\bf 5}, 9, 1099, (2018), \url{https://doi.org/10.1364/OPTICA.5.001099}

\bibitem{Liao}
 L. Liao et al., ``Proposal for an analog Schwarzschild black hole in condensates of light", Phys. Rev. A {\bf 99}, 023850, (2019), \url{https://doi.org/10.1103/PhysRevA.99.023850}
 
 \bibitem{Liu}
H. Liu, H. Guo, and R. Ling, ``Quasinormal modes of analog rotating black holes in a two-dimensional photon-fluid model", 
Phys. Rev. D {\bf 110}, 024035, (2024), \url{https://doi.org/10.1103/PhysRevD.110.024035}

\bibitem{Borsevici}
V. Borsevici, ``Condensed light, quantum black holes and L-CDM cosmology: Experimentally suggested and tested unified approach to dark matter, dark energy, cosmogenesis and two-stage inflation", {\it The Sixteenth Marcel Grossmann Meeting}, pp. 2672-2691 (2023), \url{https://doi.org/10.1142/9789811269776_0215}



%\bibitem{Mukhanov}
%V. Mukhanov, ``Evaporation and Entropy of Quaantized Black Hole", Lebedev Institute preprint N163, (1986), \url{https://inspirehep.net/literature/233701} 
%\bibitem{Mukhanov1}
%V. Mukhanov, ``Quantum Black Holes", arXiv:1810.03525, (2018) \url{https://doi.org/10.48550/arXiv.1810.03525}

\bibitem{Bekenstein4}
Jacob D.Bekenstein, Quantum Black Holes as Atoms. (1997). \url{https://doi.org/10.48550/ARXIV.GR-QC/9710076}.

\bibitem{Hooft}
G. 't Hooft, ``A planar diagram theory for strong interactions", Nucl Phys. B {\bf 72},  461, (1974), \url{https://doi.org/10.1016/0550-3213(74)90154-0}

\bibitem{Susskind}
L. Susskind, ``The World as a Hologram", Jour. Math. Phys., {\bf 36}, 11, 6377–6396, (1995), \url{https://doi.org/10.1063/1.531249}

\bibitem{Witten}
E. Witten, ``Anti De Sitter Space And Holography", Adv. Theor. Math. Phys., {\bf 2}, 2, 253–291, (1998), \url{https://doi.org/10.4310/ATMP.1998.v2.n2.a2}

\bibitem{Mukhanov2}
J. D. Bekenstein and V. F. Mukhanov, ``Spectroscopy of the quantum black hole", Phys. Lett. B {\bf 360}, 7-12, (1995), \url{https://doi.org/10.1016/0370-2693(95)01148-J}

\bibitem{Unruh}
W. G. Unruh, ``Notes on black-hole evaporation", Phys. Rev. D {\bf 14}, 870, (1976), \url{https://doi.org/10.1103/PhysRevD.14.870}

\bibitem{Jacobson}
T. Jacobson, ``Thermodynamics of Spacetime: The Einstein Equation of State", Phys. Rev. Lett. {\bf 75}, 1260, (1995), \url{https://doi.org/10.1103/PhysRevLett.75.1260}



%\bibitem{Agullo}
%I. Agullo, et al., ``Toward the Observation of Entangled Pairs in BEC Analogue Expanding Universes",  arXiv:2411.09596, (2024),
\url{ https://doi.org/10.48550/arXiv.2411.09596}



\bibitem{Hawking0}
S. W. Hawking, ``Black hole explosions?", Nature, {\bf 248}, 30–31, (1974),\url{ https://doi.org/10.1038/248030a0}

\bibitem{LoPresto}
M. C. LoPresto, ``Some Simple Black Hole Thermodynamics", Phys. Teach. {\bf 41}, 299–301 (2003), \url{https://doi.org/10.1119/1.1571268}


\bibitem{Barrow}
J. D. Barrow and B. J. Carr, ``Formation and evaporation of PBHs in scalar-tensor gravity theories", Phys. Rev. D, {\bf 54}, 6, 3920, (1996), \url{https://doi.org/10.1103/PhysRevD.54.3920}

\bibitem{Tinyakov}
P. Tinyakov, ``PBHs: the asteroid mass window", 	arXiv:2406.03114, (2024), \url{https://doi.org/10.48550/arXiv.2406.03114}

\bibitem{Carr1}
B. J. Carr, ``The Primordial Black Hole Mass Spectrum" Astrophys. Jour., {\bf 201}, 1–19,  (1975), \url{https://ui.adsabs.harvard.edu/link_gateway/1975ApJ...201....1C/doi:10.1086/153853}

\bibitem{Preskill}
J. Preskill, ``Do Black Holes Destroy Information?", 	arXiv:hep-th/9209058, (1992), \url{https://doi.org/10.48550/arXiv.hep-th/9209058} 

\bibitem{Susskind1}
L. Susskind, {\it The Cosmic Landscape: String Theory and the Illusion of Intelligent Design},  Hachette Book Group
 237 Park Avenue, New York, NY 10017, (2006)
\bibitem{Shannon}
C.E. Shannon, ``A Mathematical Theory of Communication", Bell Sys. Tech. Jour., {\bf 27}, 379-423, (1948),
\url{http://dx.doi.org/10.1002/j.1538-7305.1948.tb01338.x}
\bibitem{Dirac}
P. A. M. Dirac, {\it The Principles of Quantum Mechanics}, 3rd Edition, Oxford, The Clarendon Press, (1948)

\bibitem{Hawkings3}
Hawking, S. W. ‘Gravitational Radiation from Colliding Black Holes’. Physical Review Letters, {\bf 26}, 21, (1971), \url{https://doi.org/10.1103/PhysRevLett.26.1344.}

\bibitem{Isi}
I. Maximiliano et al. ``Testing the Black-Hole Area Law with GW150914’. Phys. Rev. Lett., {\bf 127}, 1, (2021),\url{https://doi.org/10.1103/PhysRevLett.127.011103}

\bibitem{Anderson}
P. W. Anderson, ``Four Last Conjectures", 	arXiv:1804.11186, (2018), \url{https://doi.org/10.48550/arXiv.1804.11186}

\bibitem{Wheeler1}
 J. A. Wheeler,  {\it Geometrodynamics and the Issue of the Final State},  Gordon and Breach, (1964)
\bibitem{Hooft1}
G. 't Hooft, ``Dimensional Reduction in Quantum Gravity", 	arXiv:gr-qc/9310026, (1993), \url{https://doi.org/10.48550/arXiv.gr-qc/9310026}
\bibitem{Rovelli}
C. Rovelli and L. Smolin, ``Discreteness of area and volume in quantum gravity", Nucl.Phys. B {\bf 442}, 593-622, (1995); Erratum-ibid. B {\bf 456}, 753, (1995), \url{https://doi.org/10.1016/0550-3213%2895%2900150-Q} 
\bibitem{Bojowald}
M. Bojowald, ``Loop Quantum Cosmology", Living Rev. Relativ. {\bf 8}, 11, (2005), \url{https://doi.org/10.12942/lrr-2005-11}

\bibitem{Schroeder}
G. Schroeder, {\it Genesis and the Big Bang Theory}, Bantam Books, New York, (1990)

\bibitem{Weinberg}
S. Weinberg, ``The cosmological constant problem", Rev. Mod. Phys. {\bf 61}, 1, (1989), \url{https://doi.org/10.1103/RevModPhys.61.1}
\bibitem{Peebles}
P. J. E. Peebles and B. Ratra, ``The cosmological constant and dark energy", Rev. Mod. Phys. {\bf 75}, 559, (2003), \url{https://doi.org/10.1103/RevModPhys.75.559
}
\bibitem{Padmanabhan}
T. Padmanabhan, ``Cosmological constant—the weight of the vacuum", Phys. Rep. {\bf 380}, 5–6, 235-320, (2003), \url{https://doi.org/10.1016/S0370-1573(03)00120-0}
\bibitem{Kiefer}
C. Kiefer, {\it Quantum Gravity}, 3rd Ed., Int. Series Monog. Phys., Oxford University Press, Oxford, (2012), \url{https://global.oup.com/academic/product/quantum-gravity-9780199585205?cc=us&lang=en&#}
\bibitem{Carroll}
S. M. Carroll, ``The Cosmological Constant" Living Rev. Relativ. {\bf 4}, 1 (2001), \url{https://doi.org/10.12942/lrr-2001-1}
\bibitem{Yoo}
J. Yoo and Y. Watanabe, ``Theoretical models of dark energy", Int. J. Mod. Phys. D {\bf 21}, 1230002, (2012), \url{https://doi.org/10.1142/S0218271812300029}
\bibitem{Panda}
A. Panda et al., `` Cosmological effects on $f(R,T)$ gravity through a non-standard theory", Int. J. Mod. Phys. D, {\bf 33}, 3 \& 4, 2450015, (2024), \url{https://dx.doi.org/10.1142/S0218271824500159}


\bibitem{Aghanim}
N. Aghanim et al., ``Planck 2018 results. VI. Cosmological parameters",  
Astron. \& Astrophys., {\bf 641}, A6, 67, (2020), \url{https://ui.adsabs.harvard.edu/link_gateway/2020A&A...641A...6P/doi:10.1051/0004-6361/201833910}

\bibitem{Riess}
A. G. Riess et al., ``Large Magellanic Cloud Cepheid Standards Provide a $1\%$ Foundation for the Determination of the Hubble Constant and Stronger Evidence for Physics beyond $\Lambda$CDM", Astrophys. Jour., {\bf 876}, 85, (2019), \url{https://doi.org/10.3847/1538-4357/ab1422}




\bibitem{Weinberg1}
S. Weinberg, {\it Cosmology}, Cambridge University Press, Cambridge (2008).

\bibitem{Dodelson}
S. Dodelson, {\it Modern Cosmology}, Academic Press, (2003).

\bibitem{Kolb}
E. W. Kolb and M. S. Turner, {\it The Early Universe}, Addison-Wesley, (1990).

\bibitem{Carroll1}
S. M. Carroll, {\it Spacetime and Geometry: An Introduction to General Relativity}, Cambridge University Press, Cambridge, (2019)





\bibitem{Vaidya} P.C. Vaidya, ``The gravitational field of a radiating star" Proc. Indian Acad. Sci. Sect. A,  {\bf 33}, 264, (1951), \url{https://doi.org/10.1007/BF03173260}
\bibitem{Husain}
V. Husain, ``Exact solutions for null fluid collapse" Phys. Rev. D, {\bf 53}, R1759(R), (1996), \url{https://doi.org/10.1103/PhysRevD.53.R1759} 
\bibitem{Manna} G. Manna, ``Gravitational collapse for the K-essence emergent Vaidya spacetime", Eur. Phys. J. C, {\bf 80}, 813, (2020) \url{https://doi.org/10.1140/epjc/s10052-020-8383-y}.


\bibitem{Carr2}
 B. J. Carr et al., ‘New Cosmological Constraints on PBHs’. Physical Review D, {\bf 81}, 10, (2010), \url{https://doi.org/10.1103/PhysRevD.81.104019}


\bibitem{Zu}
L. Zu, L. Feng, Q. Yuan,  Y. Fan,  ``Stringent constraints on the light boson model with supermassive black hole spin measurements``. Euro. Phys. J. Plus, {\bf135}, 9, (2020). \url{https://doi.org/10.1140/epjp/s13360-020-00734-9}

\bibitem{Lundgren}
B. F. Lundgren et al., ``Tracing the Mass Growth and Star Formation Rate Evolution of Massive Galaxies from $z \sim 6$ to $z \sim 1$ in the Hubble Ultra-Deep Field", Astrophys. Jour., {\bf 780}, 34, (2014), \url{http://dx.doi.org/10.1088/0004-637X/780/1/34}

\bibitem{Rubin}
V. C. Rubin and W. K. Ford (Jr.),  ``Rotation of the Andromeda Nebula from a Spectroscopic Survey of Emission Regions", Astrophys. Jour., {\bf 159}, 379-403, (1970), \url{https://doi.org/10.1086/150317}
\bibitem{Ade}
P. A. R. Ade et al., ``Planck 2013 results. XVI. Cosmological parameters", Astrophys. Astron., {\bf 571}, A16, (2014), \url{https://doi.org/10.1051/0004-6361/201321591}
\bibitem{Ferreira}
T. Ferreira et al., ``X-Ray–Cosmic-Shear Cross-Correlations: First Detection and Constraints on Baryonic Effects", Phys. Rev. Lett. {\bf 133}, 051001, (2024), \url{https://doi.org/10.1103/PhysRevLett.133.051001}

\bibitem{Dine}
M. Dine and A. Kusenko, ``Origin of the matter-antimatter asymmetry", Rev. Mod. Phys. {\bf 76}, 1, (2003), \url{https://doi.org/10.1103/RevModPhys.76.1}
\bibitem{Canetti}
L. Canetti, M. Drewes and M. Shaposhnikov, ``Matter and antimatter in the universe", New J. Phys. {\bf 14}, 095012, (2012), \url{https://doi.org/10.1088/1367-2630/14/9/095012}
\bibitem{Davidson}
S. Davidson, E. Nardi and Y. Nir, ``Leptogenesis", Phys. Rept. {\bf 466}, 105-177, (2008), \url{https://doi.org/10.1016/j.physrep.2008.06.002}

\bibitem{Mukhanov3}
V. Mukhanov, {\it ``Physical Foundations of Cosmology"}, Cambridge University Press; 2005, \url{https://scholar.google.co.in/scholar?hl=en&as_sdt=0%2C5&as_vis=1&q=V.+Mukhanov%2C+%E2%80%9CPhysical+Foundations+of+Cosmology%E2%80%9D%2C+Cambridge+Univer-+sity+Press%3B+2005%2C+%28Chapters+5+and+8%29%2C+pp%3A+237%2C+338&btnG=}

\bibitem{Weinberg2}
S. Weinberg, {\it Gravitation and Cosmology: Principles and Applications of the General Theory of Relativity}, Wiley Student Edition,
John Wiley \& Sons (Asia) Pte. Ltd., (2004)





\bibitem{Carr3}
B. Carr and J. Silk, ``Primordial black holes as generators of cosmic structures", Month. N. R. Astron. Soc., {\bf 478}, 3, 3756–3775, (2018),  \url{https://doi.org/10.1093/mnras/sty1204}

\bibitem{Bennett}
C. L. Bennett et al., ``Nine-year Wilkinson Microwave Anisotropy Probe (WMAP) Observations: Final Maps and Results", Astrophys. J. Suppl. Series, {\bf 208}, 2, 20, (2013), \url{https://doi.org/10.1088/0067-0049/208/2/20}

\bibitem{Carr4}
B. J. Carr, ``Primordial black holes as a probe of cosmology and high energy physics" Lecture Notes in Physics, {\bf 631}, 301–321, (2003), \url{https://doi.org/10.1007/978-3-540-45230-0_7}


\bibitem{Bambi}
C. Bambi and A. Santangelo editors, "Handbook of X-ray and Gamma-ray Astrophysics", Springer, 2024,
\url{https://link.springer.com/referencework/10.1007/978-981-19-6960-7}


\bibitem{Kahn}
Steven M. Kahn, "The cosmic X-Ray Backgroung", Kalvi Institute for Particle Astrophysics and Cosmology, Stanford University
\url{https://www.slac.stanford.edu/econf/C0307282/lec_notes/kahn/kahn1.pdf}


\bibitem{Hickox}
R. C. Hickox and M. Markevitch, ``Absolute Measurement of the Unresolved Cosmic X-Ray Background in the 0.5-8 keV Band with Chandra", 	Astrophys. Jour., {\bf 645}, 95-114, (2006), \url{https://doi.org/10.1086/504070}
\bibitem{Green2}
A. M. Green and B. J. Kavanagh, ``Primordial black holes as a dark matter candidate", J. Phys. G: Nucl. Part. Phys. {\bf 48}, 043001, (2021), \url{https://doi.org/10.1088/1361-6471/abc534}

\bibitem{MacGibbon}
J. H. MacGibbon and B. J. Carr, ``Cosmic rays from primordial black holes", Astrophys. Jour., {\bf 371}, 447-469, (1991), \url{https://ui.adsabs.harvard.edu/link_gateway/1991ApJ...371..447M/doi:10.1086/169909}

\bibitem{Page}
D. N. Page and  S. W. Hawking, ``Gamma rays from primordial black holes", Astrophys. Jour., {\bf 206}, 1-7, (1976), \url{https://ui.adsabs.harvard.edu/link_gateway/1976ApJ...206....1P/doi:10.1086/154350}

\bibitem{Cappelluti}
N. Cappelluti et al., ``The Chandra COSMOS Legacy Survey: Energy Spectrum of the Cosmic X-Ray Background and Constraints on Undetected Populations", Astrophys. Jour., {\bf 837}, 19, (2017), \url{https://doi.org/10.3847/1538-4357/aa5ea4}

\bibitem{Blok}
W. J. G. de Blok, ``The Core-Cusp Problem", Advan. Astron., {\bf 2010}, 789293, (2009), \url{https://doi.org/10.1155/2010/789293}

\bibitem{Ferrara}
A. Ferrara, A. Pallottini and P. Dayal, , ``On the stunning abundance of super-early, luminous galaxies revealed by JWST" Month. N. R. Astron. Soc., {\bf 522}, 3, 3986–3991, (2023), \url{https://doi.org/10.1093/mnras/stad1095}

\bibitem{Markevitch}
M. Markevitch et al.,  ‘A Textbook Example of a Bow Shock in the Merging Galaxy Cluster 1E 0657-56’, Astrophys. Jour., {\bf 567}, 1, L27, (2002) ,\url{https://doi.org/10.1086/339619}



\bibitem{Alcubierre}
M. Alcubierre and B. Bruegmann, "Simple excision of a black hole in 3+1 numerical relativity", Phys. Rev. D {\bf 63}, 104006, (2001), 	
\url{https://doi.org/10.1103/PhysRevD.63.104006}



%\bibitem{pse21aug2024}
%"Exact meaning of the mass M in the Kerr metric event horizon", Physics Stack Exchange, \url{physics.stackexchange.com}, 21 aug 2024;

%\bibitem{pse27mar2019}
%"For a given mass, how big can a Kerr black hole get?",  Physics Stack Exchange, \url{physics.stackexchange.com}, 27 mar 2019.



%\bibitem{Sarkar}
%S. Sarkar, Inference, {\bf 6}, 4, (2022),
%\url{https://inference-review.com/article/heart-of-darkness}

%\bibitem{Krasinski}
%A. Krasiński, {\it Inhomogeneous Cosmological 
%Models}, Cambridge, UK, Cambridge University Press, (1997). \url{https://doi.org/10.1017/CBO9780511721694}

%\bibitem{Peebles2}
%P. J. E. Peebles, ``Anomalies in physical cosmology", Ann. Phys. {\bf 447},  169159, (2022), %\url{https://doi.org/10.1016/j.aop.2022.169159}

%\bibitem{Secrest}
%N. J. Secrest et al., ``A Challenge to the Standard Cosmological Model", Astrophys. Jour. Lett., {\bf L31}, 937, (2022) %\url{https://iopscience.iop.org/article/10.3847/2041-8213/ac88c0/meta}

%\bibitem{Keenan}
%R. C. Keenan, A. J. Barger, and L. L. Cowie, ``Evidence for $A \sim 300$ Megaparsec Scale Under-density in the Local Galaxy Distribution",  Astrophys. Jour. Lett., {\bf 775}, 62, (2013) %\url{http://dx.doi.org/10.1088/0004-637X/775/1/62}

%\bibitem{Celerier}
%Marie-Noëlle Célérier, ``Some clarifications about Lemaître-Tolman models of the Universe used to deal with the dark energy problem",	Astronomy. Astrophys., {\bf 543},  A71, (2012),
%\url{https://doi.org/10.1051/0004-6361/201219104}

%\bibitem{Celerier2}
%Marie-Noëlle Célérier, "",	Astronom. Astrophys., {\bf 543},  A71, (2012),
%\url{https://doi.org/10.1051/0004-6361/201219104}



\end{thebibliography}
\end{document}